\documentclass[superscriptaddress,aps,showpacs,nofootinbib,twocolumn]{revtex4}
\usepackage{graphicx}
\usepackage{amsmath,amssymb}

\def\AJ{{ Ap. J.} }
\def\AJL{{ Ap. J. Lett.} }

\def\CQG{{ Class. Quantum Gravity} }

\def\FP{{ Fortschr. Physik} }
\def\GRG{{ Gen. Relativity and Gravitation} }

\def\IJMP{{ Int. J. Mod. Phys.} }

\def\MPL{{ Mod. Phys. Lett.} }
\def\MNRAS{{ Mon. Not. R. Ast. Soc.} }

\def\NP{{ Nucl. Phys.} }
\def\PL{{ Phys. Lett.} }
\def\PR{{ Phys. Rev.} }
\def\PRL{{ Phys. Rev. Lett.} }

\def\frac#1#2{{\textstyle{{#1}\over {#2}}}}

\def\lsim{\mathrel{\rlap{\lower4pt\hbox{\hskip1pt$\sim$}}
    \raise1pt\hbox{$<$}}}
\def\gsim{\mathrel{\rlap{\lower4pt\hbox{\hskip1pt$\sim$}}
    \raise1pt\hbox{$>$}}}
\def\sqr#1#2{{\vcenter{\vbox{\hrule height.#2pt
         \hbox{\vrule width.#2pt height#1pt \kern#1pt
         \vrule width.#2pt}
         \hrule height.#2pt}}}}

\def\beq{\begin{equation}}
\def\eeq{\end{equation}}
\def\beqa{\begin{eqnarray}}
\def\eeqa{\end{eqnarray}}

\begin{document}

\title{Supernovae constraints on models of dark energy reexamined}
\author{M. C. Bento}
\email{bento@sirius.ist.utl.pt} \affiliation{Departamento de
F\'{\i}sica and Centro de F\'{\i}sica Te\'orica de
Part\'{\i}culas, Instituto Superior T\'{e}cnico, Avenida Rovisco
Pais, 1049-001 Lisboa, Portugal}

\author{O. Bertolami}
\email{orfeu@cosmos.ist.utl.pt} \affiliation{
Departamento de F\'{\i}sica, Instituto Superior T\'{e}cnico,
Avenida Rovisco Pais, 1049-001 Lisboa, Portugal}

\author{N. M. C. Santos}
\email{ncsantos@cftp.ist.utl.pt} \affiliation{Departamento de
F\'{\i}sica and Centro de F\'{\i}sica Te\'orica de
Part\'{\i}culas, Instituto Superior T\'{e}cnico, Avenida Rovisco
Pais, 1049-001 Lisboa, Portugal}

\author{A. A. Sen}
\email{anjan@cftp.ist.utl.pt}
\affiliation{Departamento de F\'{\i}sica and Centro
de F\'{\i}sica Te\'orica de Part\'{\i}culas, Instituto Superior
T\'{e}cnico, Avenida Rovisco Pais, 1049-001 Lisboa, Portugal}

\begin{abstract}
We use the Type Ia Supernova gold sample data of Riess {\it et al}
in order to constrain three models of dark energy. We study the
Cardassian model, the Dvali-Turner gravity modified model and the
generalized Chaplygin gas model of dark energy - dark matter
unification. In our best fit analysis for these three dark energy
proposals we consider flat model and the non-flat model priors. We
also discuss the degeneracy of the models with the XCDM model
through the computation of the so-called jerk parameter.

\end{abstract}

\pacs{98.80.-k, 98.80.Es \hspace{61mm} Preprint DF/IST-8.2004}

\maketitle

\section{Introduction}

The surprising discovery of the present late-time acceleration of
the Universe \cite{observations} and the related  fact that most
of its energy is in the form of a mysterious dark energy is
possibly one of the most puzzling issues of modern cosmology.
Several scenarios have been put forward as a possible explanation.
A positive cosmological constant, although the simplest candidate,
is not particularly attractive given the extreme fine tuning that
is required to account for the observed accelerated expansion.
This fact has led to models where the dark energy component varies
with time, such as quintessence models \footnote{An evolving
vacuum energy was discussed somewhat earlier, see e.g.
Refs. \cite{Bronstein1933}.}. In these models, the required
negative pressure is achieved trough the dynamics of a single
(light) scalar field \cite{1field} or, in some cases, two coupled
scalar fields \cite{2field}. Despite some pleasing features, these
models are not entirely satisfactory, since in order to achieve
$\Omega_X \sim \Omega_{m}$ (where $\Omega_X$ and $\Omega_{m}$ are
the dark energy and matter energy densities at present,
respectively) some fine tuning is required. Many other
possibilities have been considered for the origin of this dark
energy component  such as a scalar field with a non-standard
kinetic term and k-essence models \cite{kessence}; it is also
possible to construct models which have $w_X = p/\rho < -1$, the
so-called phantom energy models \cite{phantom}.

Recently, it has been proposed that the evidence for a dark energy
component might be explained by a change in the equation of state
of the background fluid, with an exotic equation of state, the
generalized Chaplygin gas (GCG) model
\cite{Kamenshchik:2001cp,Bilic:2001cg,Bento:2002ps}. The striking
feature of this model is that it allows for an unification of dark
energy and dark matter \cite{Bento:2002ps}.

Another possible explanation for the accelerated expansion of the
Universe could be the infrared modification of  gravity one should
expect from, for instance, extra dimensional physics, which would
lead to a modification of the effective Friedmann equation at late
times \cite{Dvali:2003rk,Dvali:2000hr,Deffayet}. An interesting
variation of this proposal has been suggested by Dvali and Turner
\cite{Dvali:2003rk} (hereafter referred to as DT model). Another
possibility, also originally motivated by extra dimensions
physics, is the modification of the Friedmann equation by the
introduction of an additional nonlinear term proportional to
$\rho^n$, the so-called Cardassian model \cite{Freese:2002sq}.

Currently type Ia supernovae (SNe Ia) observations provide the
most direct way to probe the dark energy component at low to
medium redshifts. This is due to the fact that supernova data
allows for a direct measurent of the luminosity distance, which  is
directly related to the expansion law of the Universe which, in
turn, is the physical quantity that is directly related with  dark
energy or that is modified by extra dimensional physics. This
approach has been explored by various groups in order to obtain
insight into the nature of dark energy.  Indeed, recently
supernova data with 194 data points has been analysed
\cite{Choudhury:2003tj} and it was shown that it yields relevant
constraints on some cosmological parameters. In particular, it is
possible to conclude that, when one considers the full supernova
data set, the decelerating model is ruled out with a significant
confidence level. It is also shown that one can measure the
current value of the dark energy equation of state with higher
accuracy and the data prefers the phantom kind of equation of
state, $w_X < -1$. Furthermore, the most significant result of
that analysis is that, without a flat prior, that supernova data
does not favor a flat $\Lambda$CDM model at least up to
$68\%$ confidence level,
which is consistent with other cosmological observations. In what concerns
the equation of state of the dark energy component,  it has been
shown in Ref. \cite{Alam:2003}, using the same set of
supernovae data, that  the best fit equation of state of dark energy evolves
rapidly from $w_X \simeq 0$ in the past to $w_X \lsim -1$ in the
present, which suggests that a time varying dark energy is better
fitted with the data than the  $\Lambda$CDM model. This result is also
robust to changes of $\Omega_{m}$ and remains valid for the interval
$0.1\leq \Omega_{m} \leq 0.5$.
Supernova data has also been used in the context of different
cosmological models for dark energy
\cite{Bertolami2004,Perivol2004,Gong:2004sa,Zhu:2004ij,Elgaroy:2004ne,Sen:2003cy}.
In this paper, we analyze the Cardassian, the DT and the GCG
models in light of the Riess {\it et al.} SNe Ia compilation of
data \cite{Riess:2004nr}. We consider both flat and non-flat
priors.

Notice however, that our analysis is restricted to the very
late history of the Universe and does not
address dark energy effects on the cosmic microwave
background fluctuations or on structure formation.

This paper is organized as follows. In section II we describe our best
fit analysis of SNe Ia data that will be employed as methodology to
constrain the dark energy models we have studied, namely the
Cardassian, DT and GCG models.  In section III we consider
the best fit analysis to constrain the Cardassiam dark energy model.
In section IV, we discusss the DT model and in section V the dark
energy - dark matter unification GCG model.  Section VI is devoted to
the discussion of the degeneracy of the discussed models with the XCDM
model through the introduction of the so-called jerk parameter. In
section VII we present our conclusions.

\section{Observational constraints from supernovae data}

The observations of supernovae measure essentially the
apparent magnitude $m$, which is related to the luminosity distance $d_L$ by
\begin{align}
m(z) = {\cal M} + 5 \log_{10} D_L(z) ~,
\end{align}
where
\begin{align}
D_L(z) \equiv {{H_0}\over{c}} d_L(z)~, \label{DL}
\end{align}
is the dimensionless luminosity distance and
\begin{align}
d_L(z)=(1 + z) d_M(z)~,
\label{dL}
\end{align}
with $d_M(z)$ being the comoving distance given by
\begin{align}
d_M(z)=c \int_0^z {{1}\over{H(z')}} dz'~. \label{dm}
\end{align}
Also,
\begin{align}
{\cal M} = M + 5 \log_{10}
\left({{c/H_0}\over{1~\mbox{Mpc}}}\right) + 25~,
\end{align}
where $M$ is the absolute magnitude which is believed to be constant for
all supernovae of type Ia.

For our analysis, we consider the two sets of supernovae data
recently compiled  by Riess {\it et al.} \cite{Riess:2004nr}. The
first set contains 143 points from previously published data that
were taken from the 230 Tonry {\it et al.} \cite{Tonry:2003zg}
data alongwith the 23 points from Barris {\it et al.}
\cite{Barris:2003dq}. They have discarded various points where the
classification of the supernovae was not certain or the photometry
was incomplete, increasing the reliability of the sample. The
second set contains the 143 points from the first one plus 14
points discovered recently using Hubble Space Telescope (HST)
\cite{Riess:2004nr} and is named as gold sample, following Riess
{\it et al.}. We will name the first sample as gold w/o HST. The
main difference between the two samples is that the full gold
sample also covers higher redshifts ($1<z<1.6$).

The data points in these samples are given in terms of the distance modulus
\begin{align}
\mu_{\rm obs}(z) \equiv m(z) - M_{\rm obs}(z)~,
\end{align}
and the errors $\sigma_{\mu_{\rm obs}}(z)$ already quoted take into
 account the effects of peculiar motions.

The $\chi^2$ is calculated from
\begin{align}
\chi^2 = \sum_{i=1}^n \left[ {{\mu_{\rm obs}(z_i) - {\cal M}' - 5
\log_{10}D_{L \rm th}(z_i; c_{\alpha})}\over{\sigma_{\mu_{\rm
obs}}(z_i)}} \right]^2~,
 \label{chisq2}
\end{align}
where ${\cal M}' = {\cal M} - M_{\rm obs} $ is a free parameter
and $D_{L \rm th}(z; c_{\alpha})$ is the theoretical prediction
for the dimensionless luminosity distance of a supernova at a
particular distance, for a given model with parameters
$c_{\alpha}$. It can be computed for each model from the Friedmann
expansion law (cf. Eqs. (\ref{FriedCmodel2}), (\ref{FriedDTmodel})
or (\ref{FriedGCGmodel}) below) combined with Eqs.
(\ref{DL}-\ref{dm}).

In the following, we will consider the three models referred to in
the Introduction and perform a best fit analysis with the
minimization of the $\chi^2$, Eq. (\ref{chisq2}), with respect to
${\cal M}'$, $\Omega_{m}$, $\Omega_k$  and the respective model
parameter(s), using a MINUIT \cite{Minuit} based code.

The allowed variation range of the parameters is presented in
Table \ref{table:range}. ${\cal M}'$ is a model independent
parameter and hence its best fit value should not depend on the
specific model. We found that the best fit value for ${\cal M}'$
for all the models considered here is $43.3$, which is consistent
with the one obtained in \cite{Choudhury:2003tj}. Hence, we have
also used this value for ${\cal M}'$ throughout our analysis. We
have also checked that the result does not change  if we
marginalise over $\cal{M}'$.

\begin{table}[t]
\begin{tabular}{c c}
 \hline \hline
\hspace{5mm} Parameters \hspace{5mm} & \hspace{5mm} Range \hspace*{5mm}\\
 \hline
$\Omega_{m}$ or $A_s$ & $]0,1[$\\

$\Omega_{k}$ & $]-1,1[$\\

${\cal M}'$ & $[41,45]$\\
 \hline
Cardassian: $n$ & $[-30,2/3]$\\

DT: $\beta$ & $[-60,1]$\\

GCG: $\alpha$ & $[0,10]$\\
 \hline \hline
\end{tabular}
\caption{Parameter range for the best-fit analysis}
\label{table:range}
\end{table}

In what follows we shall present a description of the models that are
considered in this analysis and perform the best fit study considering flat and
non-flat priors. Our results are summarized in Table \ref{table:best}.

\begin{table*}[t!]
\begin{tabular}{c c c c c c}
 \hline \hline
  & \hspace*{4mm} Data Sample \hspace*{4mm}  & $\Omega_{m}$ or $A_s$
  & Model
 parameter
& $\Omega_{k}$ & $\chi^2$ \\
 \hline
 {\bf Cardassian model} & & \hspace*{22mm} & & \hspace*{22mm} &
  \hspace*{22mm} \\
\hline
 Flat &  Gold w/o HST & $0.53$ & $-2.0$ & $-$ & $154.6$ \\
 Prior &  Gold  & $0.49$ & $-1.4$ & $-$ & $173.7$ \\

 Non-Flat &  Gold w/o HST & $0.97$ & $-0.93$ & $-0.75$ & $154.4$ \\
 Prior &  Gold & $0.21$ & $-3.1$ & $0.47$ & $173.2$ \\
\hline
 {\bf \hspace{-15mm} DT model} & &  & & & \\
 \hline
 Flat &  Gold w/o HST & $0.55$ & $-60.0$ & $-$ & $155.4$ \\
 Prior &  Gold  & $0.51$ & $-19.2$ & $-$ & $174.7$ \\
 \hline
 Non-Flat &  Gold w/o HST & $1.0$ & $-15.6$ & $-0.71$ & $155.0$ \\
 Prior &  Gold & $0.24$ & $-60.0$ & $0.43$ & $174.0$ \\
\hline
 {\bf \hspace{-12mm} GCG model} & &  & & & \\
 \hline
 Flat &  Gold w/o HST & $0.98$ & $6.2$ & $-$ & $155.0$ \\
 Prior &  Gold  & $0.93$ & $2.8$ & $-$ & $174.2$ \\
  \hline
 Non-Flat &  Gold w/o HST & $0.73$ & $1.3$ & $-1.0$ & $154.2$ \\
 Prior &  Gold & $0.97$ & $4.0$ & $0.02$ & $174.5$ \\
 \hline\hline
\end{tabular}
\caption{Best fit parameters for the Cardassian, DT and GCG
models, for the two data samples of Riess {\it et al.}, considering
flat and non-flat priors. The best fit value used for ${\cal M}'$
is $43.3$.} \label{table:best}
\end{table*}

\begin{figure*}[htb!]
\begin{center}
 \includegraphics[height=6.5cm]{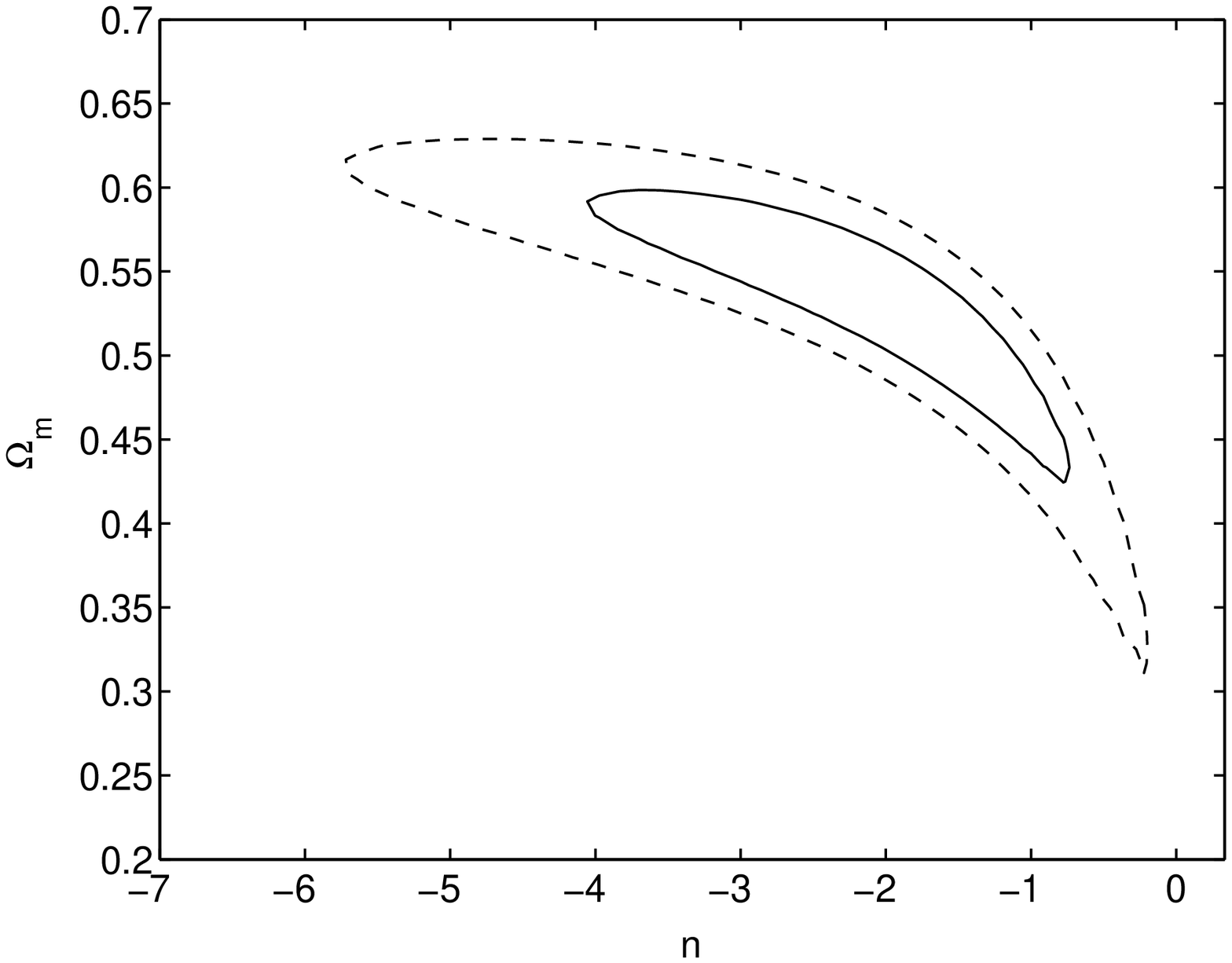}
 \includegraphics[height=6.5cm]{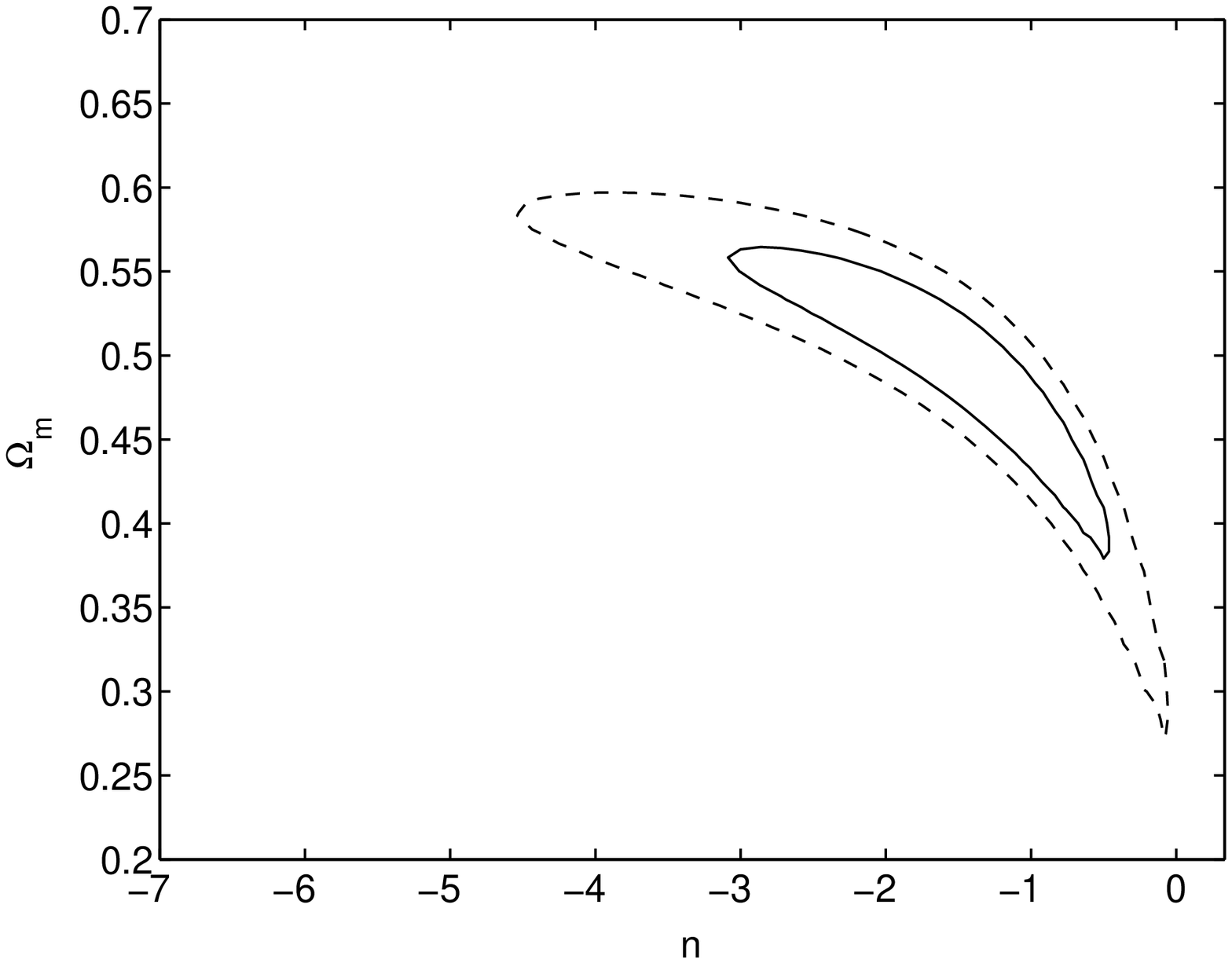}
  \caption{\label{fig:flatcard} Confidence contours in the
  $\Omega_{m}-n$ parameter space for the flat Cardassian model. The
  solid and dashed lines represent the $68\%$ and $95\%$ confidence
  regions, respectively; in the left panel are shown the results for
  the gold sample without the HST SNe Ia, whereas in the right panel, the full
  gold sample is taken into account.}
\end{center}
\end{figure*}

\section{Cardassian model}

We first consider the so Cardassian model \cite{Freese:2002sq}, which
justifies the late time accelerating universe by invoking an
additional term in the Friedmann equation proportional to $\rho^n$.
In this model, the universe is composed only of radiation and matter
(including baryon and cold dark matter) and the increasing expansion
rate is given by
\begin{align}
\label{FriedCmodel} H^2= {{8 \pi}\over {3 M_{\rm Pl}^2}} \left(
\rho + b \rho^n \right) - {{k} \over {a^2}}
\end{align}
where $M_{\rm Pl} = 1.22 \times 10^{19}$GeV is the $4$-dimensional
Planck mass, $b$ and $n$ are constants, and we have added a
curvature term to the original Cardassian model. At present, the
universe is matter dominated, i.e. $\rho_{m} \gg \rho_{rad}$,
hence $\rho \approx \rho_{m}$. The new term dominates only
recently, at about $z \sim 1$, hence in order to get the  recent acceleration
in the expansion rate, $n < 2/3$ is required.

\begin{figure*}[htb!]
\begin{center}
 \includegraphics[height=6.5cm]{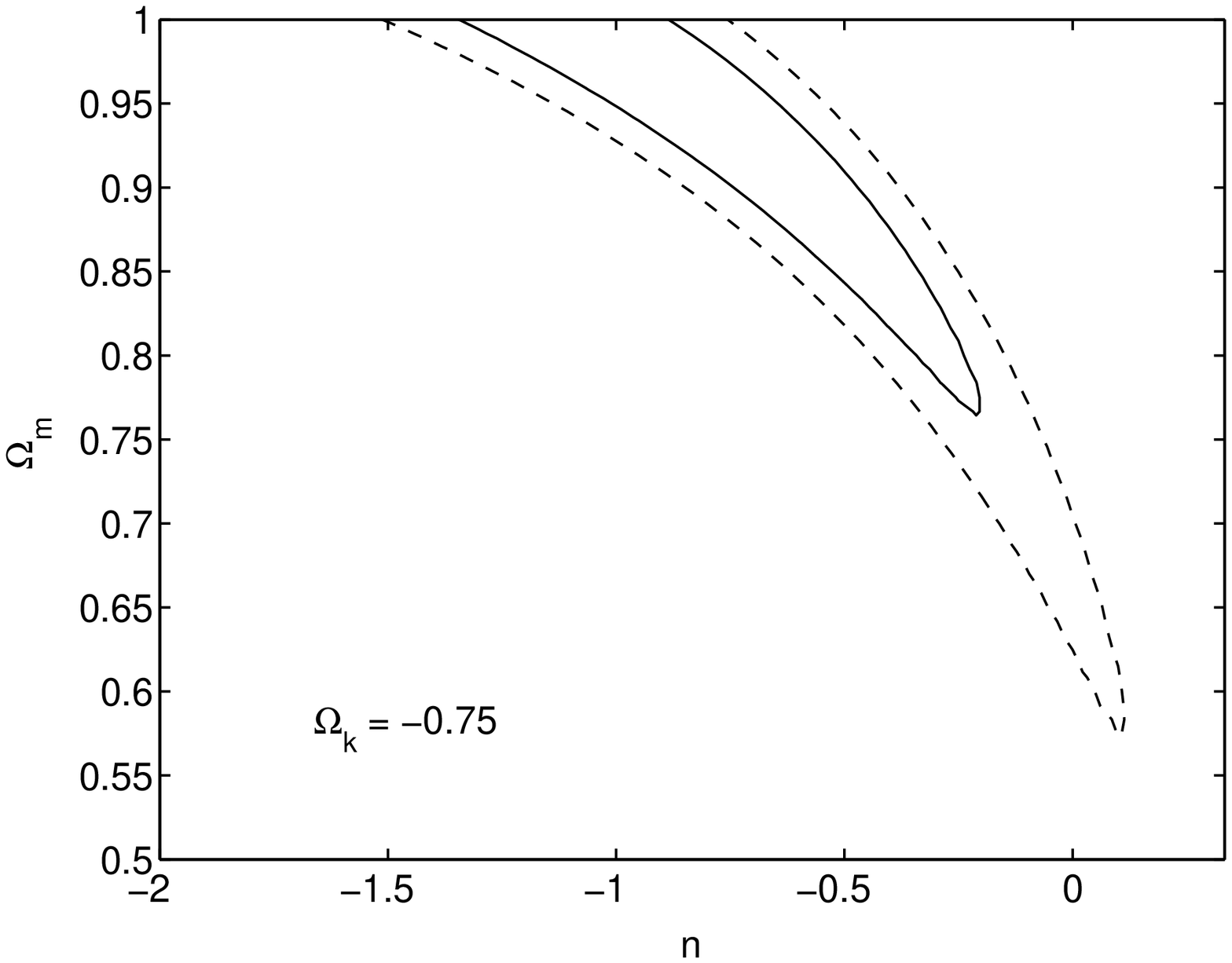}
 \includegraphics[height=6.5cm]{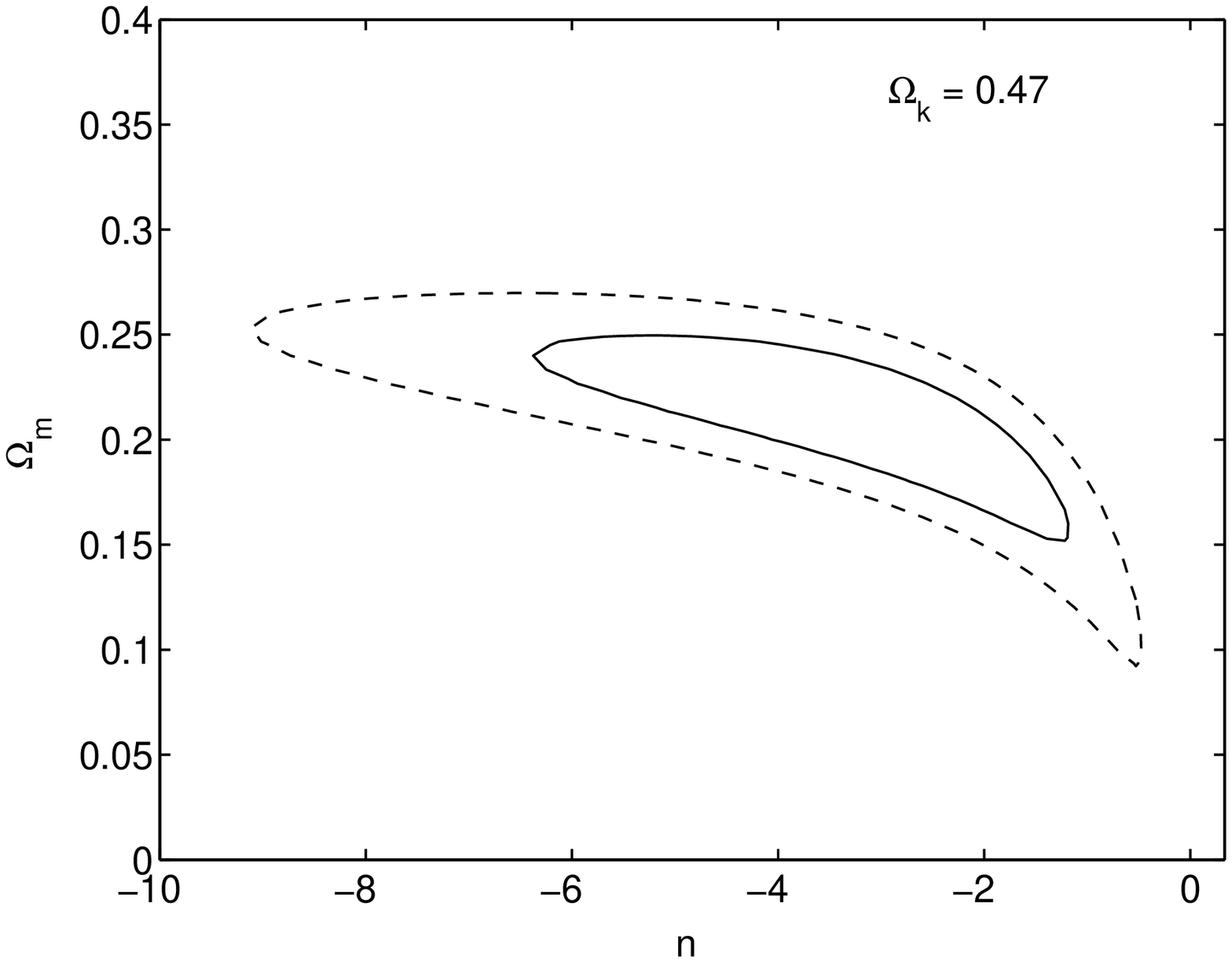}
\includegraphics[height=6.5cm]{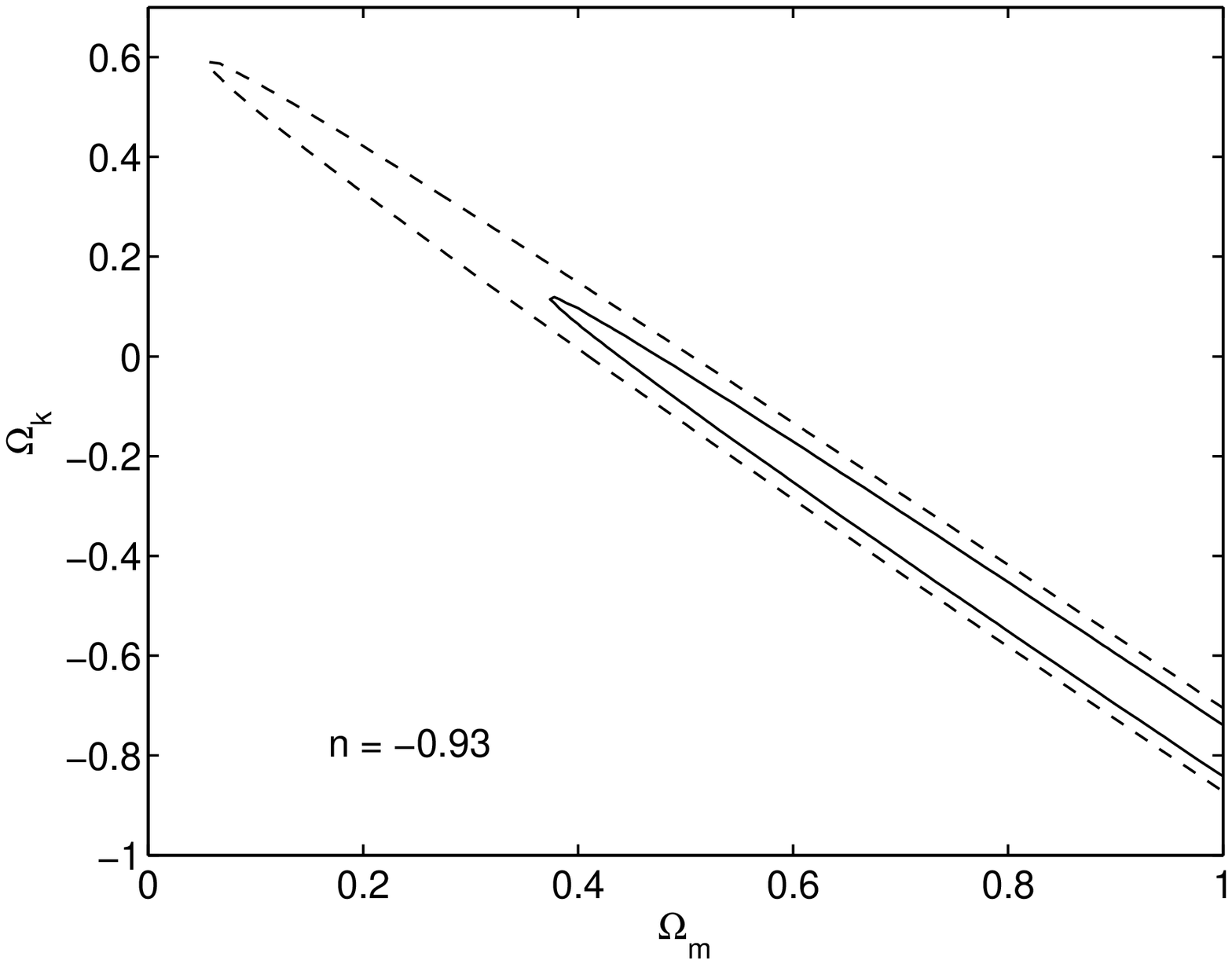}
 \includegraphics[height=6.5cm]{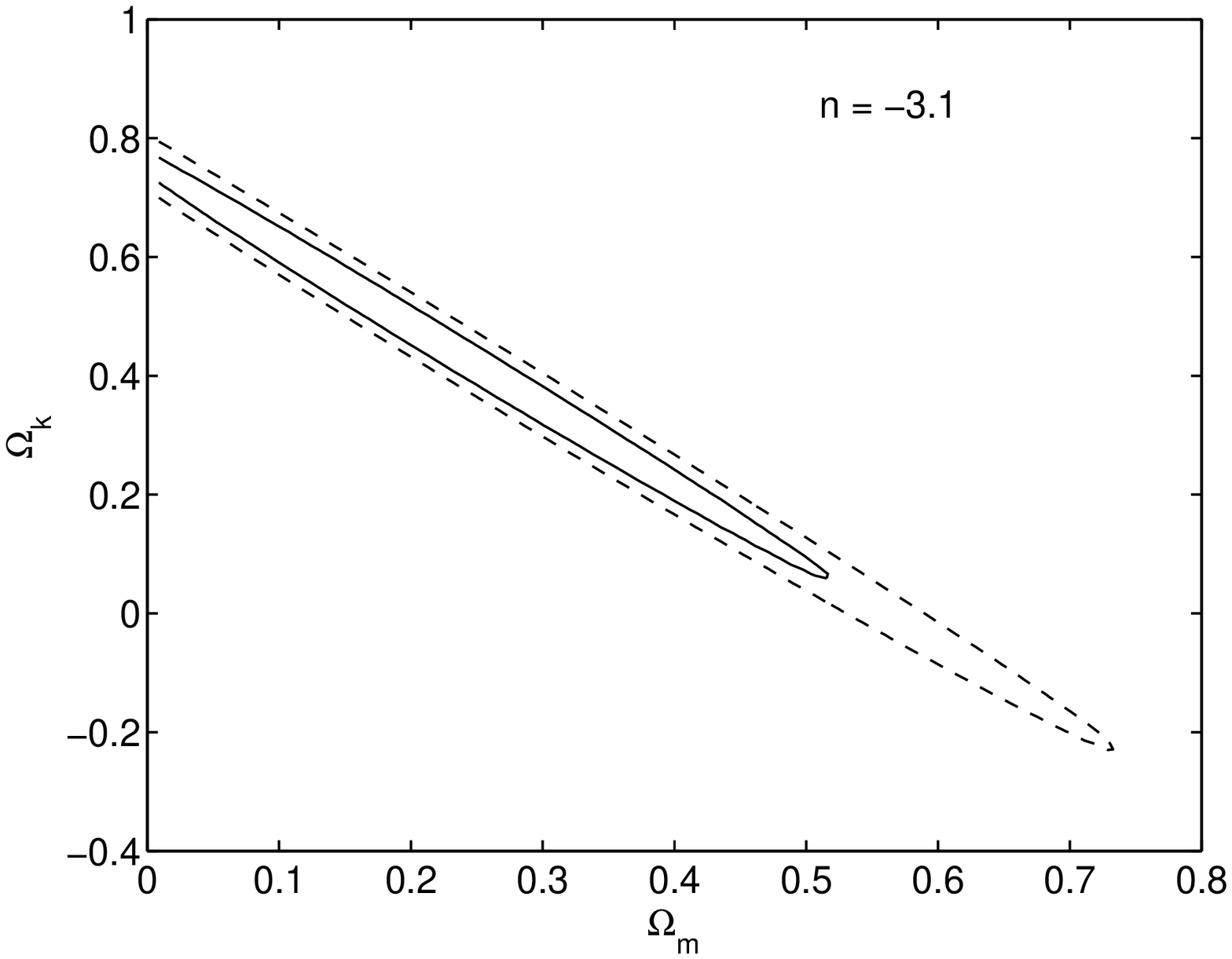}
\includegraphics[height=6.5cm]{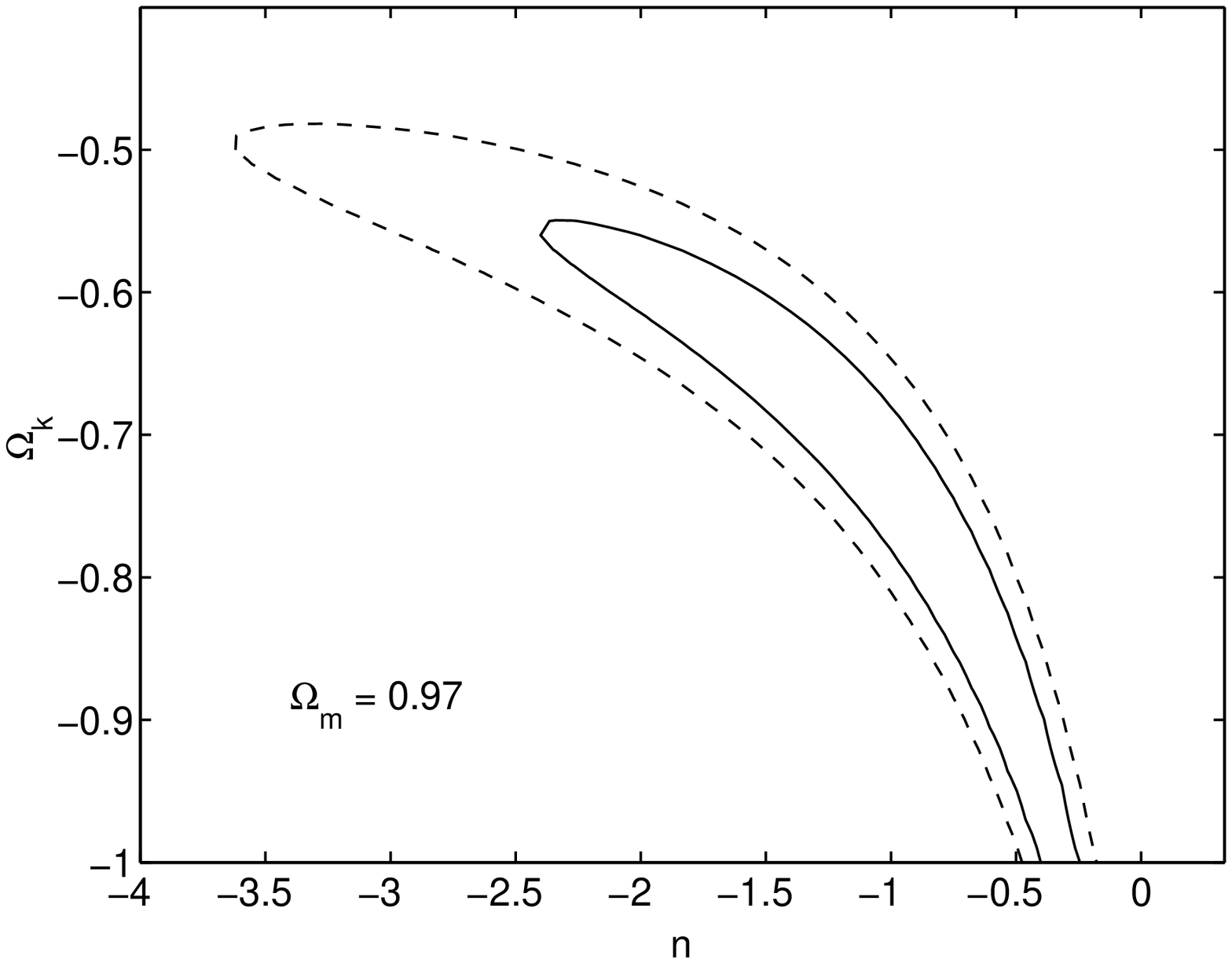}
 \includegraphics[height=6.5cm]{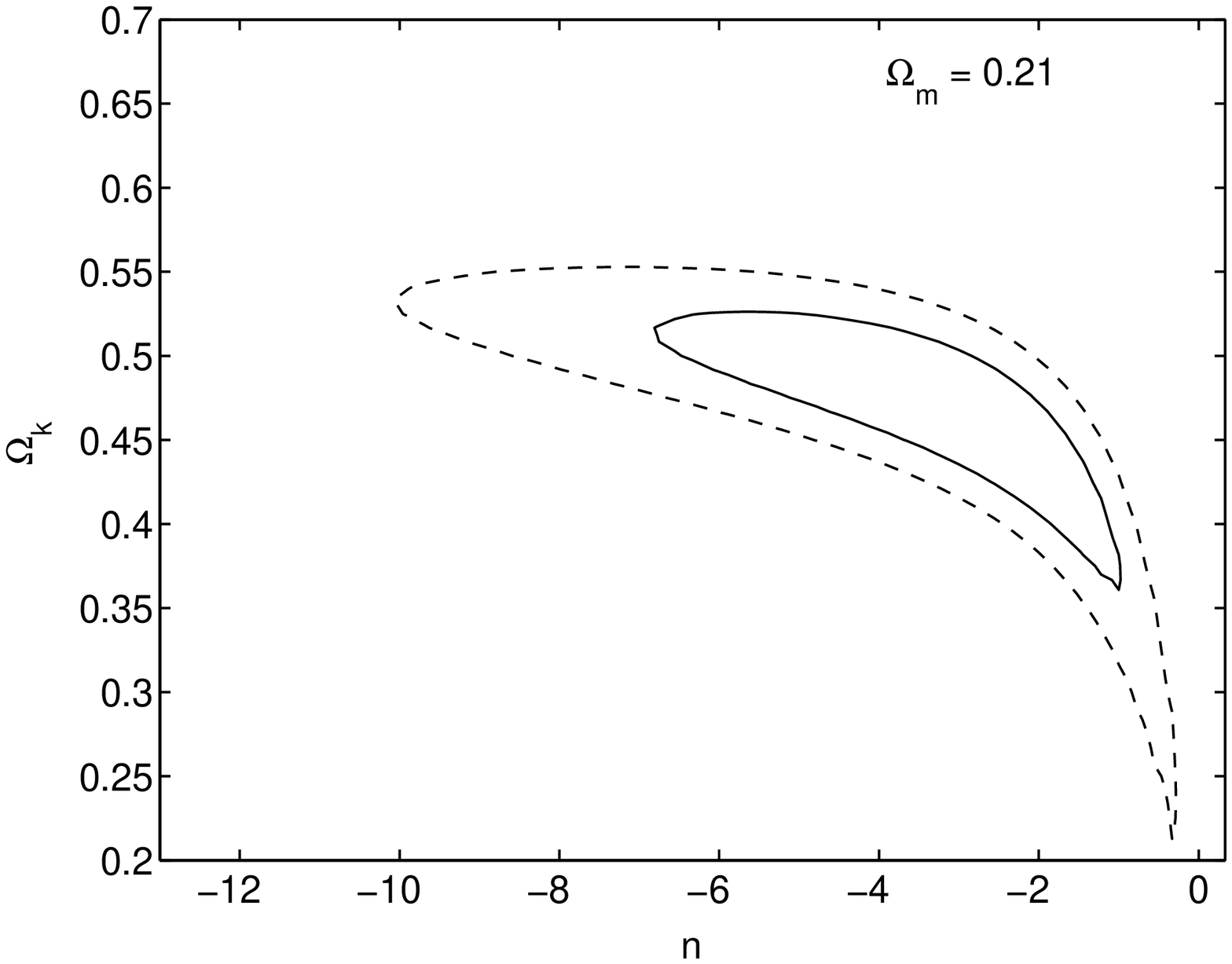}
\caption{\label{fig:card} Confidence contours in the and
$\Omega_{m}-n$, $\Omega_{k}-\Omega_{m}$ and $\Omega_{k}-n$
parameter space
  for the non-flat Cardassian model. As in Figure \ref{fig:flatcard}, the
  solid and dashed lines represent
  the $68\%$ and $95\%$ confidence regions, respectively; in the left
  panel is used the golden sample without HST SNe Ia, whereas in the right
  one the full gold sample is taken into account.}
\end{center}
\end{figure*}

\begin{figure*}[htb!]
\begin{center}
 \includegraphics[height=6.5cm]{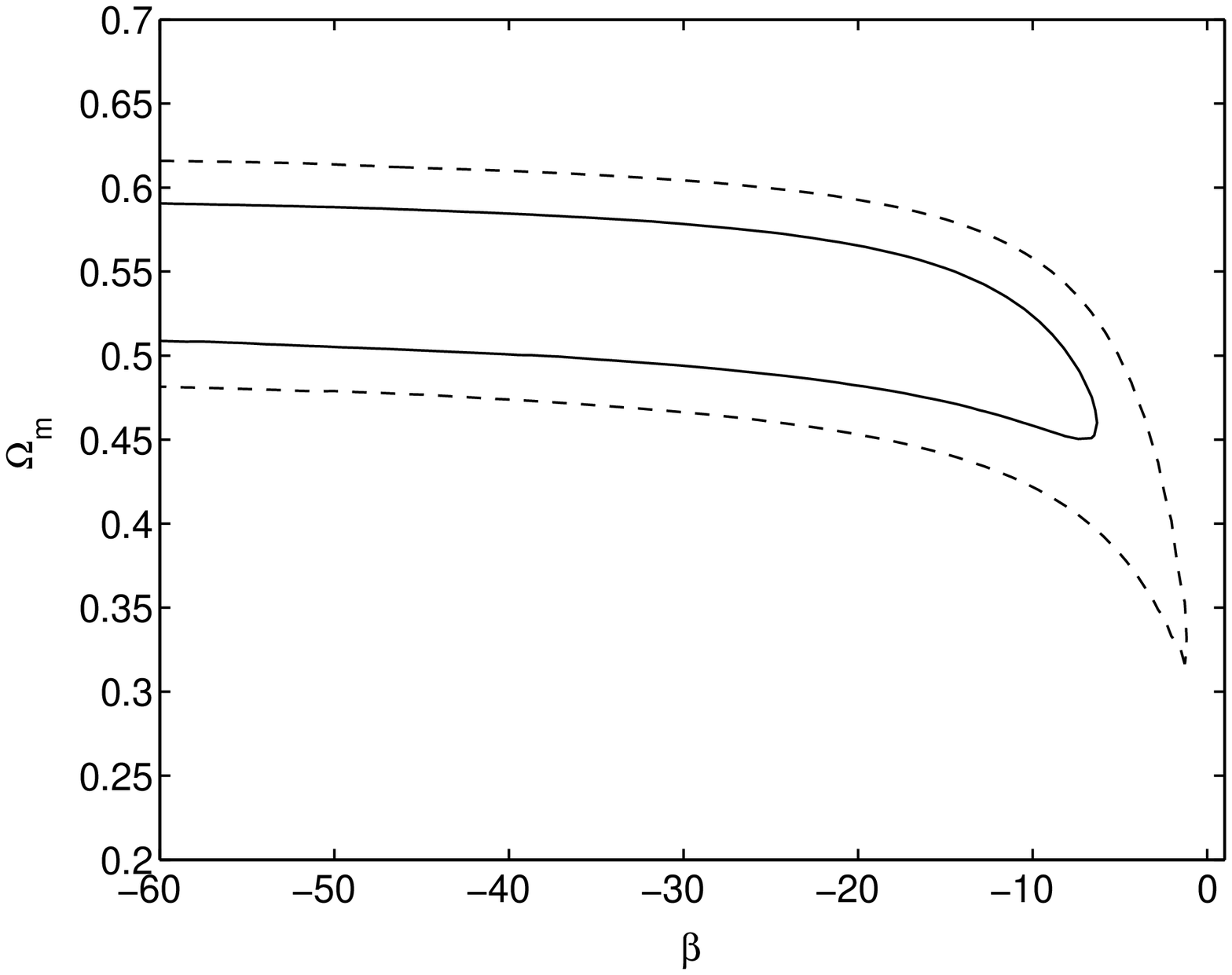}
 \includegraphics[height=6.5cm]{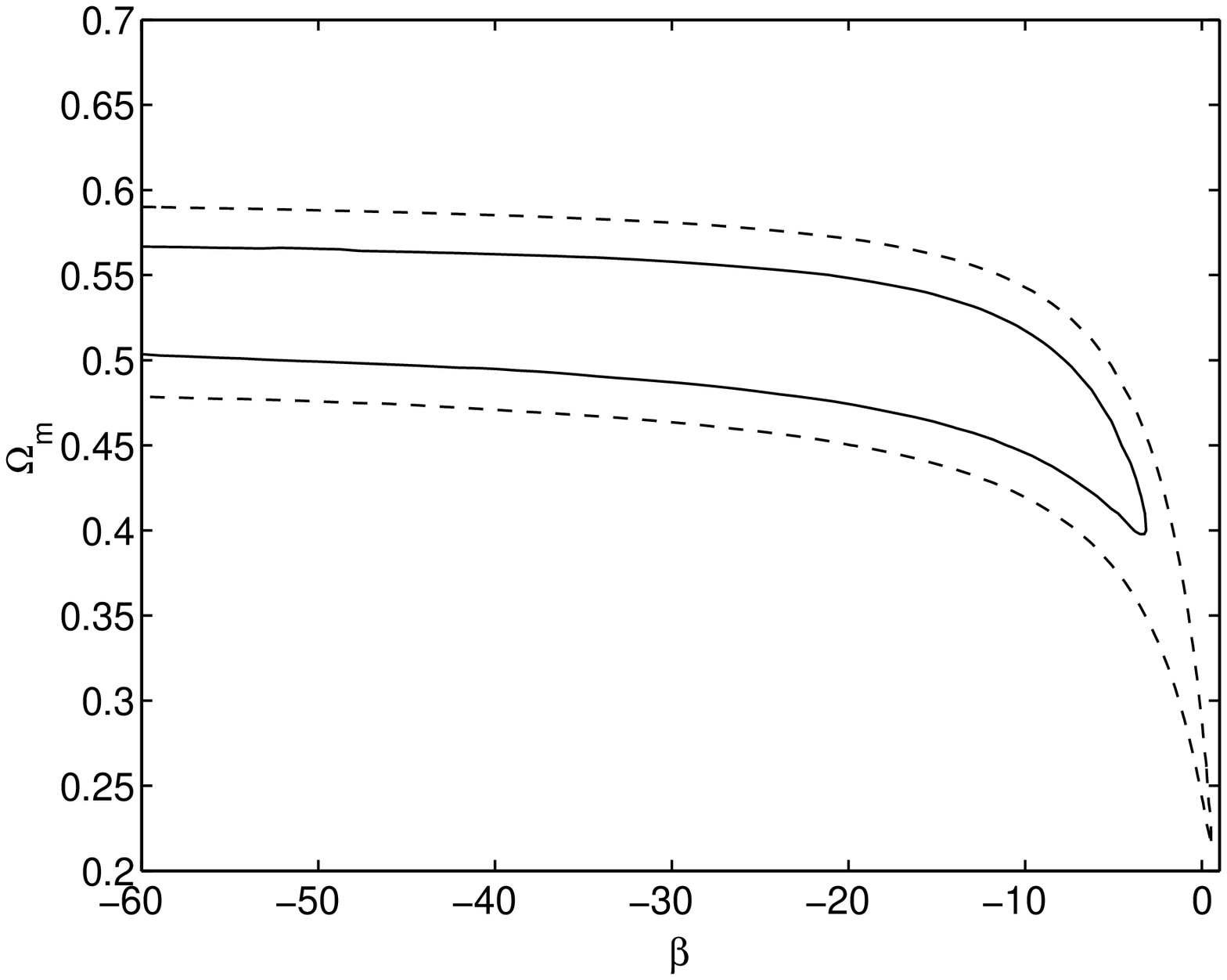}
  \caption{\label{fig:flatDT} Confidence contours in the
  $\Omega_{m}-\beta$ parameter space for the flat DT model. The
  solid and dashed lines represent the $68\%$ and $95\%$ confidence
  regions, respectively; in the left panel are shown the results for
  the gold sample without the HST SNe Ia, whereas in the right the full
  gold sample is taken into account.}
\end{center}
\end{figure*}

\begin{figure*}[htb!]
\begin{center}
 \includegraphics[height=6.5cm]{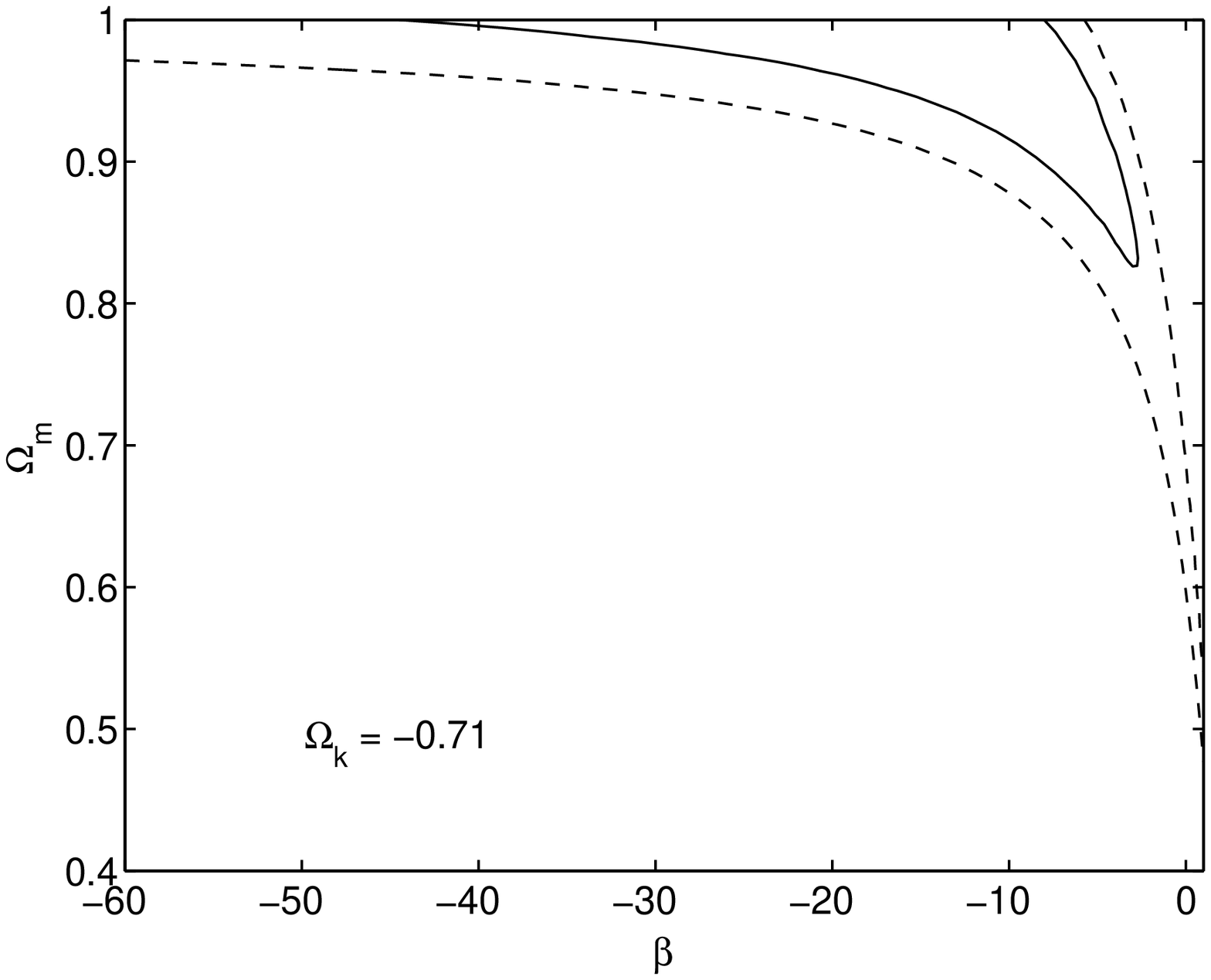}
 \includegraphics[height=6.5cm]{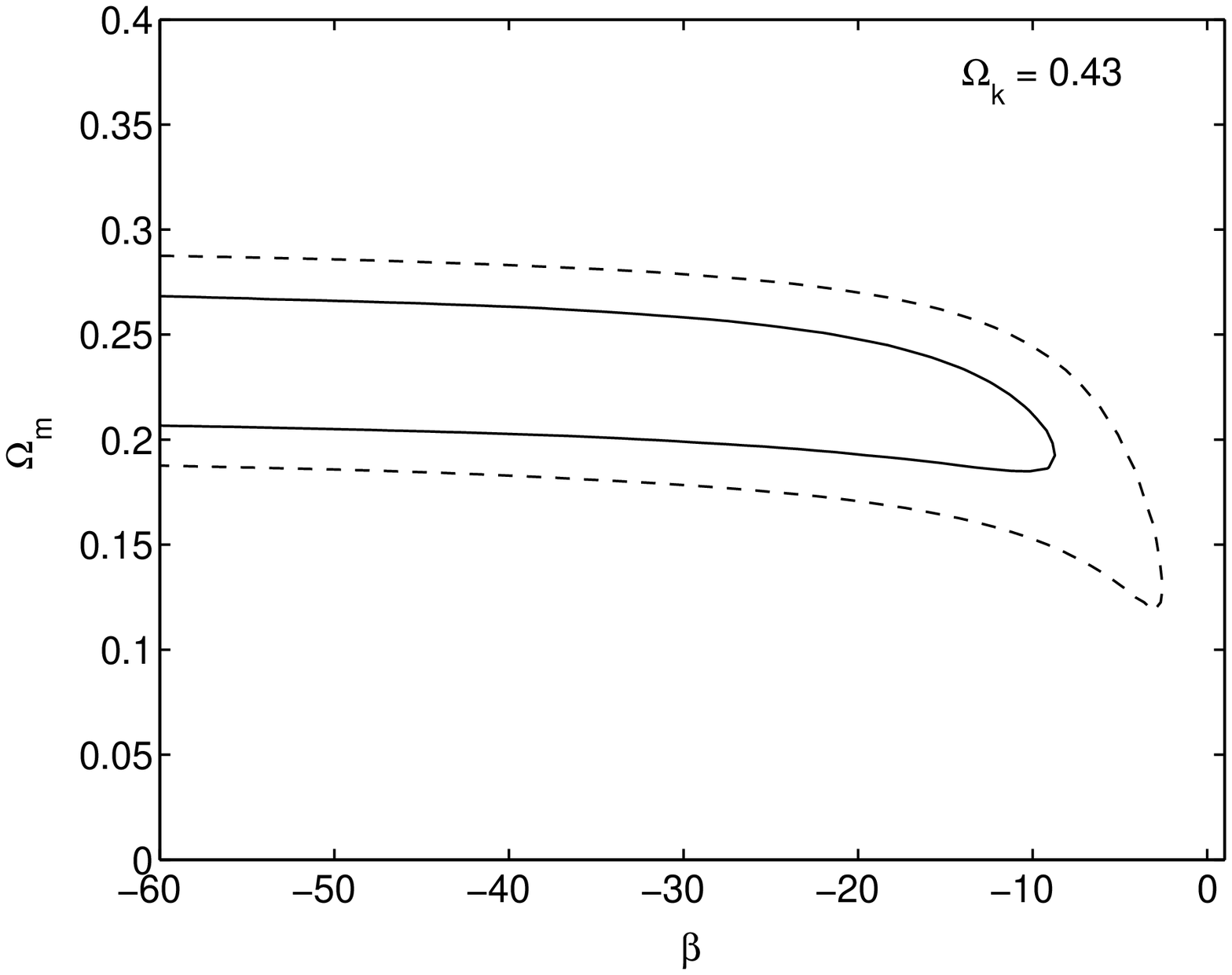}
\includegraphics[height=6.5cm]{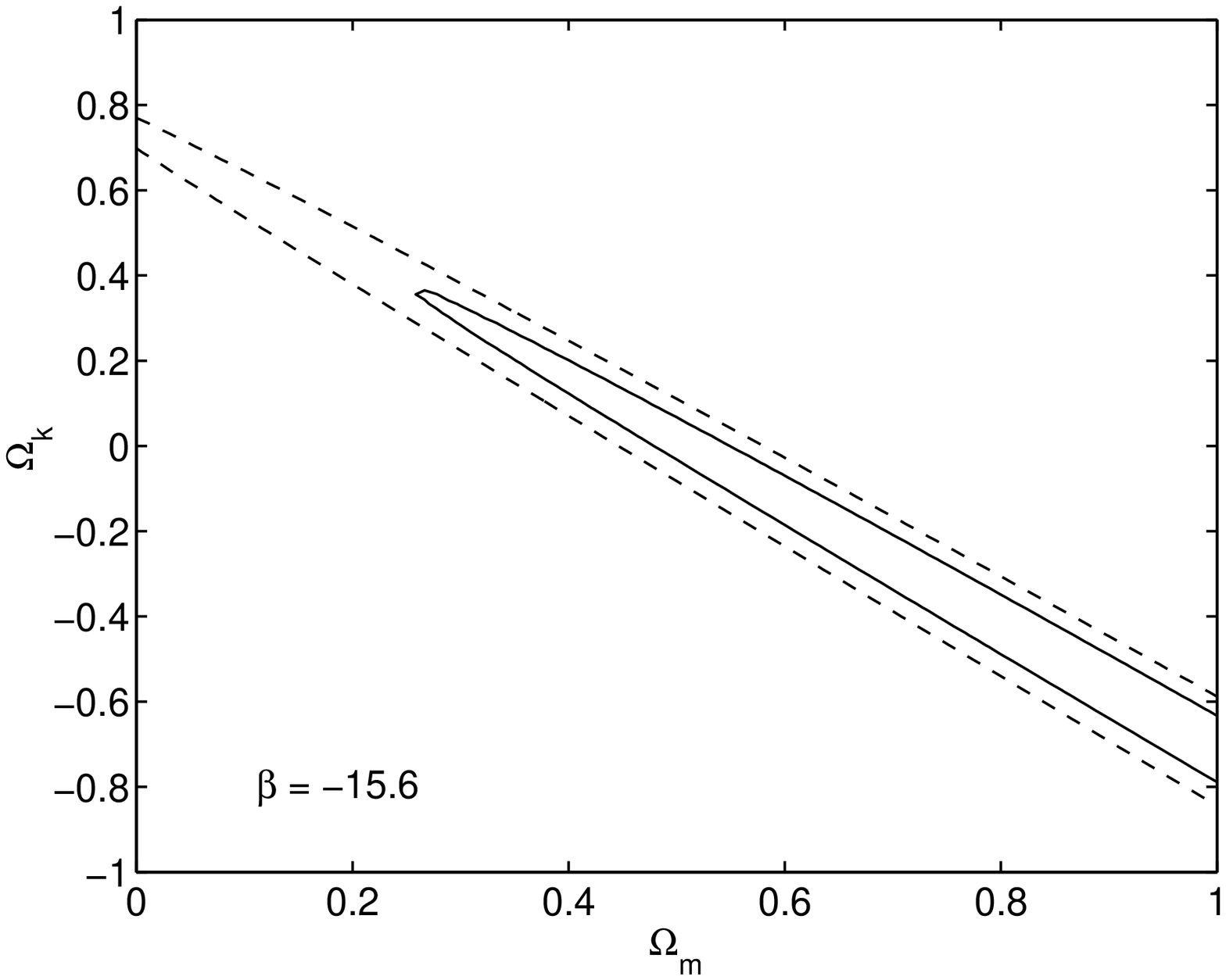}
 \includegraphics[height=6.5cm]{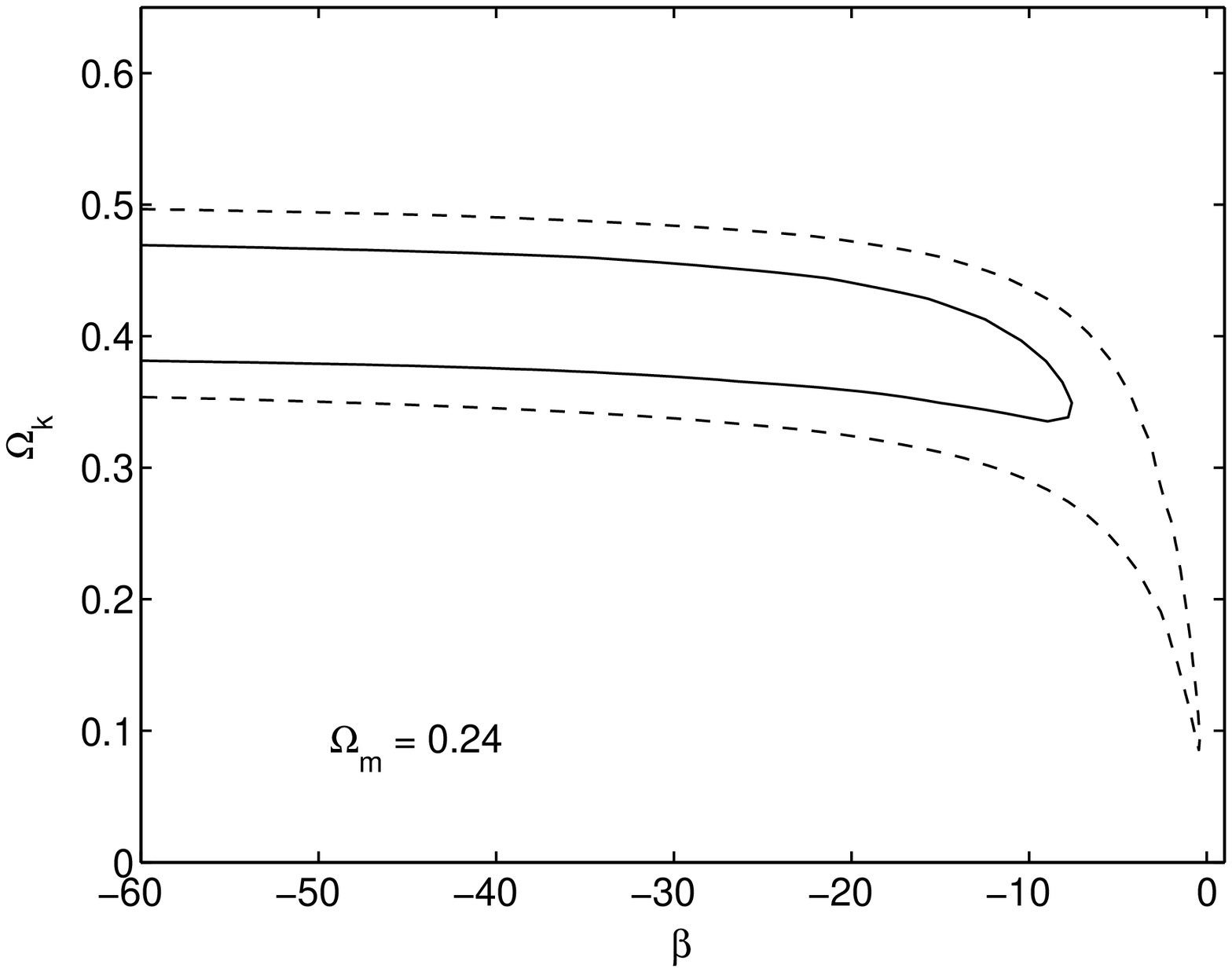}
\caption{\label{fig:DT} Confidence contours in the
$\Omega_{m}-\beta$ and $\Omega_{k}-\Omega_{m}$ parameter space
  for the non-flat DT model using the gold sample without HST
  SNe Ia (left panel) and in the $\Omega_{m}-\beta$ and
  $\Omega_{k}-\beta$ plane using the full gold sample (right
  panel). As in Figure \ref{fig:flatDT}, the
  solid and dashed lines represent
  the $68\%$ and $95\%$ confidence regions, respectively.}
\end{center}
\end{figure*}

The theoretical motivation for the Cardassian term is fairly
speculative. It can be argued that its origin may arise as a consequence
of embedding our (3+1)-dimensional brane universe in
extra-dimensions \cite{Chung:1999zs}, or from some unknown interactions
between matter particles \cite{Gondolo:2002fh}.

In a matter dominated universe,  equation (\ref{FriedCmodel})
can be rewritten as
\begin{align}
\label{FriedCmodel2} \left({{H} \over {H_0}}\right)^2= \Omega_{m}
(1+z)^3+\Omega_k (1+z)^2 \nonumber \\
+ (1-\Omega_{m}-\Omega_k)(1+z)^{3n}~,
\end{align}
where $H_0$ is the present day value of the Hubble constant and
$\Omega_k=-{{k} \over {a_0^2 H_0^2}}$ is the present curvature
parameter. Notice that  the only parameter of the model is $n$ and
that the case $n=0$ corresponds to the $\Lambda$CDM model.

The best fit results for this model, for the two data samples we are
considering, and taking into account flat
and non-flat priors, are summarized in Table \ref{table:best}.

For the flat case, we have only two parameters and the best
fitting results we get are $\{\Omega_{m},n\}=\{0.53,-2.0\}$,
without HST, and $\{\Omega_{m},n\}=\{0.49,-1.4\}$, with HST SNe
Ia. In Fig. \ref{fig:flatcard} we show the 68$\%$ and 95$\%$
confidence contour plots. These results are consistent with those
obtained in previous works by Gong $\&$ Duan \cite{Gong:2004sa},
Zhu {\it et al.} \cite{Zhu:2004ij} and Sen $\&$ Sen
\cite{Sen:2003cy} for other data sets. Hence, we see that the case $n=0$,
which corresponds to the $\Lambda$CDM model, is excluded at a 95$\%$
confidence level by both samples, even though HST sample favors
slightly larger values of $n$.

If we relax the flat prior and consider the curvature we find that
the best fit analysis reveals that the gold sample without HST
data favors a negative curvature around $\Omega_k = - 0.75$, with
a significant matter component, $\Omega_m = 0.97$, even though the
68$\%$ confidence contour region (see Fig. \ref{fig:card}) is
consistent with curvatures in the range $[-1,0.1]$ and matter
density in the range $[0.38,1]$; for the model parameter we find
that it lies in the range $[-2.4, -0.2]$ with central value
$n=-0.93$. From the 95$\%$ confidence contour, meaningful bounds
can be still obtained for the model parameter: $n$ must lie in the
range $[-3.6, 0.1]$, hence not excluding the $\Lambda$CDM model.
Values for $\Omega_k$ and, mainly, $\Omega_m$ cannot be
significantly constrained.

Clearly the HST data brings the amount of matter to lower values,
but pushes curvature to positive values and the model parameters
for values that are smaller than the ones obtained with gold
without HST data. We find that the best fitting value for the
curvature is significantly positive $\Omega_k = 0.47$ and
$\Omega_m = 0.21$; for the model parameter we obtain as best fit
value $-3.1$. Moreover, the $n=0$ case is excluded with $95\%$
confidence level. The contour plots are presented in
Fig.\ref{fig:card}.

\section{Dvali-Turner model}

The second model we will consider is the one proposed by Dvali \&
Turner \cite{Dvali:2003rk}, where the Friedmann equation is
modified by the addition of the term $H^\beta /r_c^{2-\beta}$,
which can arise in theories with extra dimensions
\cite{Dvali:2000hr}; $r_c$ is a crossover scale which sets the
scale beyond which the laws of the 4-dimensional gravity breakdown
and become 5-dimensional. In this case, in contrast with theories
with infinite volume extra dimensions, the laws of gravity are
modified in the far infrared and the cosmological evolution gets
modified at late times; the short distance gravitational dynamics
is very close to that of the 4-dimensional Einstein gravity, hence
the early times cosmological evolution is very close to the
Friedmann-Robertson-Walker picture.

As a motivation for this type of modification of gravity, consider the
model with a single extra dimension \cite{Dvali:2000hr,Deffayet}, with
the effective low-energy action given by
\begin{align}
{S}\,=\, {M_{\rm Pl}^2 \over r_c}
\,\int\,d^4x\,dy\,\sqrt{g^{(5)}}\,{\cal R}\, \nonumber\\
 +\,\int\,
d^4x \sqrt{g}\,\left ( M_{\rm Pl}^2\, R\, +\,{\cal L}_{\rm
SM}\right )\, , \label{actionDGP}
\end{align}
where $y$ is the
extra spatial coordinate, $g^{(5)}$ is the trace of the
$5$-dimensional metric $g^{(5)}_{AB}$ $(A,B=0,1,2,...,4)$, $g$ the
trace of the $4$-dimensional metric induced in the brane,
$g_{\mu\nu}(x)~\equiv~g_{\mu\nu}^{(5)}(x, y=0)$. The first term in
the action is the bulk $5$-dimensional Einstein action, where
${\cal R}$ is the $5$-dimensional Ricci scalar,  and the second one
 is the $4$-dimensional Einstein term, localized on the brane (at
$y=0$), where $R$ is the $4$-dimensional Ricci scalar and ${\cal
L}_{\rm SM}$ is the Lagrangian of the fields in the Standard
Model. For the maximally symmetric FRW ansatz, $ds_5^2 \, = \,
f(y,H) ds_4^2 \, -\, dy^2$, where $ds_4^2$ is the 4-dimensional
maximally-symmetric metric, one gets the modified Friedmann
equation on the brane
\begin{align}
\label{FriedDGPmodel}
 H^2 \, \pm \, {H \over r_c}  \, = \, {8\pi \over 3 M_{\rm Pl}^2}\rho~,
\end{align}
where $\rho$ is the total energy density in the brane.

Inspired in this construction,  Dvali \& Turner considered a more
generic (and radical) modification  of the Friedmann equation
\cite{Dvali:2003rk}
\begin{align}
H^2 \, -  \, {H^{\beta} \over r_c^{2 - \beta}} \, = \, {8\pi \over
3 M_{\rm Pl}^2} \rho - {{k} \over {a^2}}~,
\end{align}
where we have also introduced the curvature term in the brane.
The crossover scale $r_c$ is fixed in order to eliminate the need
for  dark energy,
\begin{align}
r_c = H_0^{-1}~(1- \Omega_m - \Omega_k)^{1/ (\beta - 2)}~,
\end{align}
where we assume again that the universe is matter dominated.
 The Friedmann expansion law can then be written as
\begin{align}
\label{FriedDTmodel} \left({ {H}\over{H_0}} \right)^2 = \Omega_{m}
(1+z)^3 + (1-\Omega_{m}-\Omega_k) \left({{H}\over{H_0}}
\right)^\beta \nonumber \\
+ \Omega_k (1+z)^2 ~.
\end{align}
Notice that $\beta$ is the only parameter of the model: for
$\beta=0$ the new term behaves like a cosmological constant, and
for $\beta=2$ it corresponds to a ``renormalization'' of the
Friedmann equation. Note also that the case $\beta=1$ corresponds
to the model in Eq. \ref{FriedDGPmodel}, hereafter called
Dvali-Gabadadze-Porrati (DGP) model \cite{Dvali:2000hr}. The
successful predictions of the Big-Bang nucleosynthesis impose a
limit on $\beta$, namely, $\beta \leq 1.95$; a more stringent
bound follows from requiring that the new term does not interfere
with the formation of large-scale structure: $\beta \leq 1$.
Moreover, it can be shown that, in the recent past ($10^4 > z \gg
1$), this correction behaves like dark energy with equation of
state $w_{eff}=-1+\beta/2$, and $w=-1$ in the distant future and,
moreover, it can mimic $w<-1$ without violating the weak-energy
condition \cite{Dvali:2003rk}.

For the flat DT model, the best-fitting parameters are
$\{\Omega_{m},\beta\}=\{0.55,-60.0\}$ without HST and
$\{\Omega_{m},\beta\}=\{0.51,-19.2\}$ with HST supernovae, as is
summarized in Table \ref{table:best}. From Fig. \ref{fig:flatDT},
where the 68$\%$ and 95$\%$ confidence contour plots are shown, it
is clear that $\beta$ is very weakly constrained by both
supernovae data sets, and can become arbitrarily large and
negative. Moreover, the cosmological constant, corresponding to
$\beta=0$, seems to be disfavoured; also $\beta=1$, the DGP model,
is strongly disfavoured. These results are consistent with those
of Elgaroy {\it et al.} \cite{Elgaroy:2004ne} obtained with
another data set.

If we fix $\beta=1$ (DGP model), and allow only $\Omega_{m}$ to
vary, the best fit value are $\Omega_{m} = 0.17$ and  $\Omega_{m}
= 0.16$ for the gold sample and gold sample without HST SNe Ia,
respectively, which are consistent with the results of Gong $\&$ Duan
\cite{Gong:2004sa}.

Now, if we relax the flat prior, we find that $\{\Omega_k,
\beta\}= \{-0.71, -15.6\}$ are the best fit values for the gold
sample without HST SNe Ia (Table \ref{table:best}); the best fit
value for $\Omega_m$ is in the end of the variation range we
considered for this parameter, $\Omega_m = 1$. As we can see from
the contour plots shown in Fig.\ref{fig:DT}, $\Omega_m$ cannot be
constrained at $95\%$ confidence level, however the $68\%$
confidence contours show that data prefer $\Omega_m > 0.25$. In
what concerns $\beta$, it can become arbitrarily large and
negative, if we allow $\Omega_m$ to be large. Moreover, both
$\Lambda$CDM and DGP models are disfavored with $68\%$ confidence.

Once more the HST data brings the amount of matter to lower values
and pushes curvature to positive values. The best fit results are
$\{\Omega_{m}, \Omega_k, \beta\}= \{0.24, 0.43, -60,0\}$. Also in
this case both $\Lambda$CDM and DGP models are disfavored.

\section{Generalized Chaplygin gas model}

Finally, we consider the generalized Chaplygin gas model, which is
characterized by the equation of state
\begin{align}
p_{ch} = - {A \over \rho_{ch}^\alpha}~, \label{rhoGCG}
\end{align}
where $A$ and $\alpha$ are positive constants. For $\alpha=1$, the
equation of state is reduced to the Chaplygin gas scenario
\cite{Kamenshchik:2001cp}.

Integrating the energy conservation equation with the equation of
state (\ref{rhoGCG}), one gets \cite{Bento:2002ps}
\begin{align}
\rho_{ch} = \rho_{ch0} \left[A_{s} + {(1-A_s) \over
a^{3(1+\alpha)}}\right]^{1/(1+\alpha)}~,
\end{align}
where $\rho_{ch0}$ is the present energy density of GCG and $A_s
\equiv A/\rho_{ch0}^{(1+\alpha)}$. One of the most striking
features of this expression is that the energy density of this
GCG, $\rho_{ch}$, interpolates between a dust dominated phase,
$\rho_{ch} \propto a^{-3}$, in the past and a de-Sitter phase,
$\rho_{ch} = -p_{ch}$, at late times. This property makes the GCG
model an interesting candidate for the unification of dark matter
and dark energy. Moreover,  one can see from the above equation
that $A_s$ must lie in the range $0\le A_s \le 1$: for $A_s =0$,
GCG behaves always as matter whereas for $A_s =1$, it behaves
always as a cosmological constant. We should point out, however,
that if we want to unify dark matter and dark energy, one has to
exclude these two possibilities resulting the range for $A_s$ as
$0< A_s < 1$. Notice also that $\alpha = 0$ corresponds to the
$\Lambda$CDM model.

\begin{figure*}[htb!]
\begin{center}
 \includegraphics[height=6.5cm]{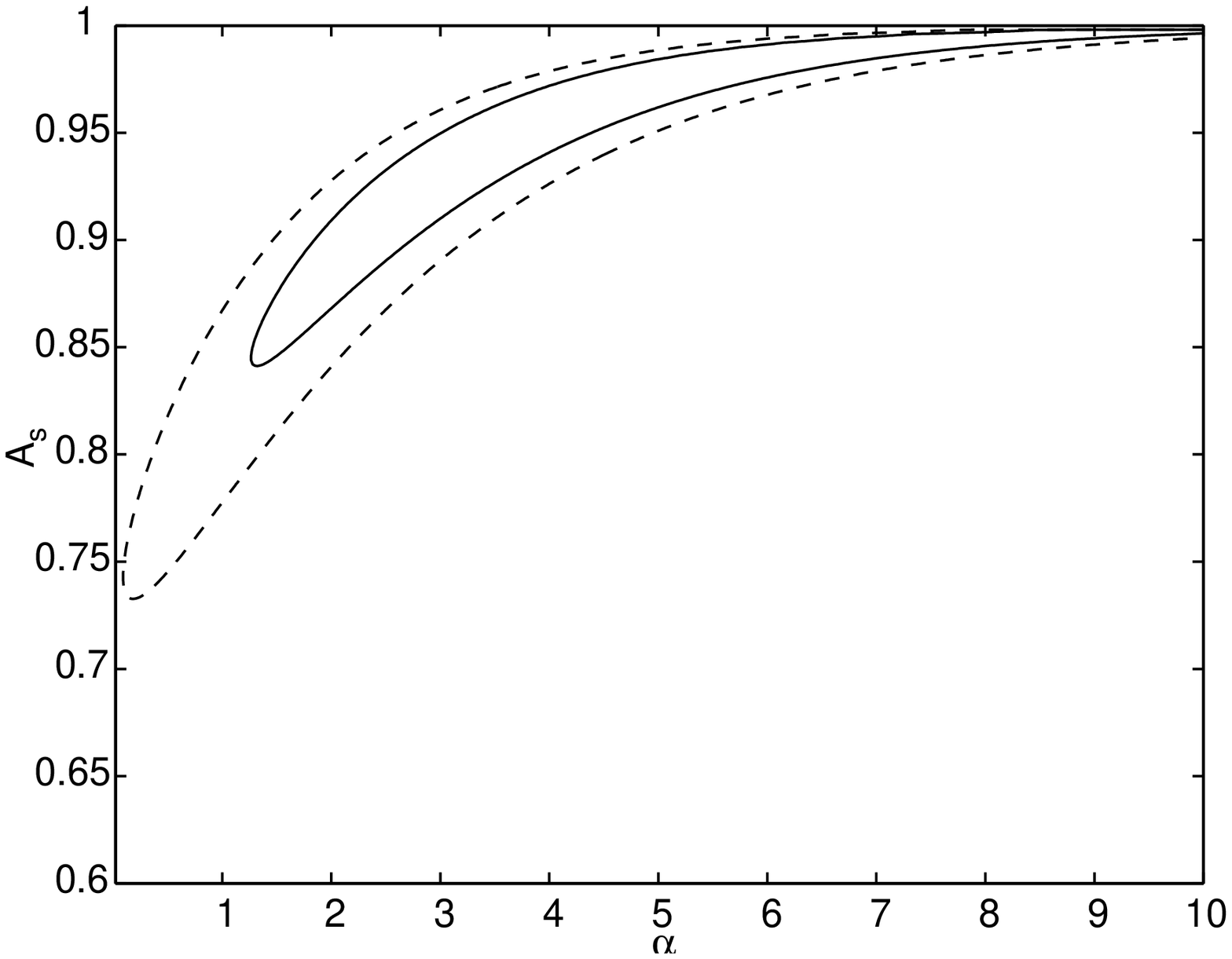}
 \includegraphics[height=6.5cm]{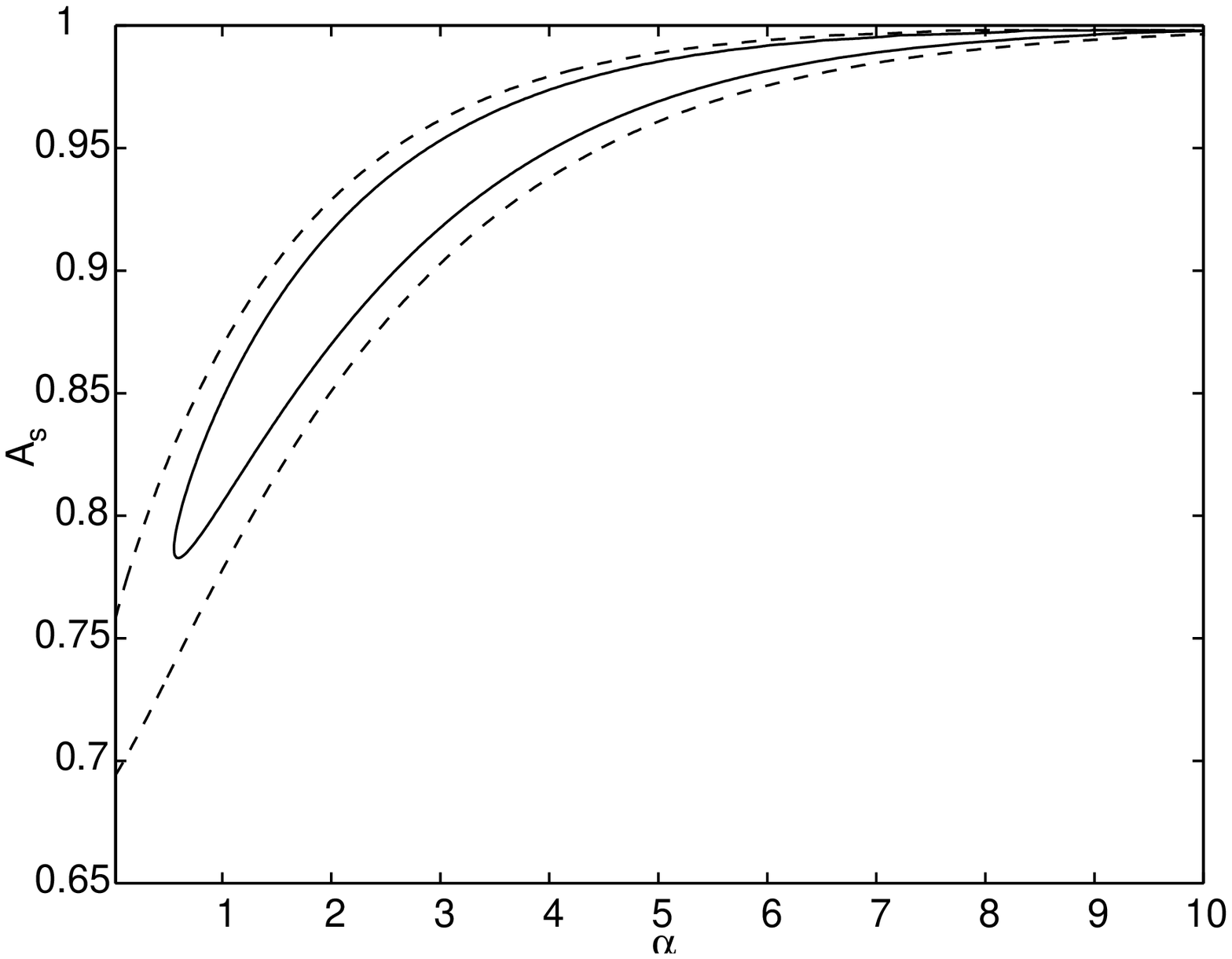}
  \caption{\label{fig:GCGflat} Confidence contours in the and $A_s-\alpha$,
parameter space for the flat GCG model. The
  solid and dashed lines represent
  the $68\%$ and $95\%$ confidence regions, respectively.}
\end{center}
\end{figure*}

\begin{figure*}[htb!]
\begin{center}
 \includegraphics[height=6.5cm]{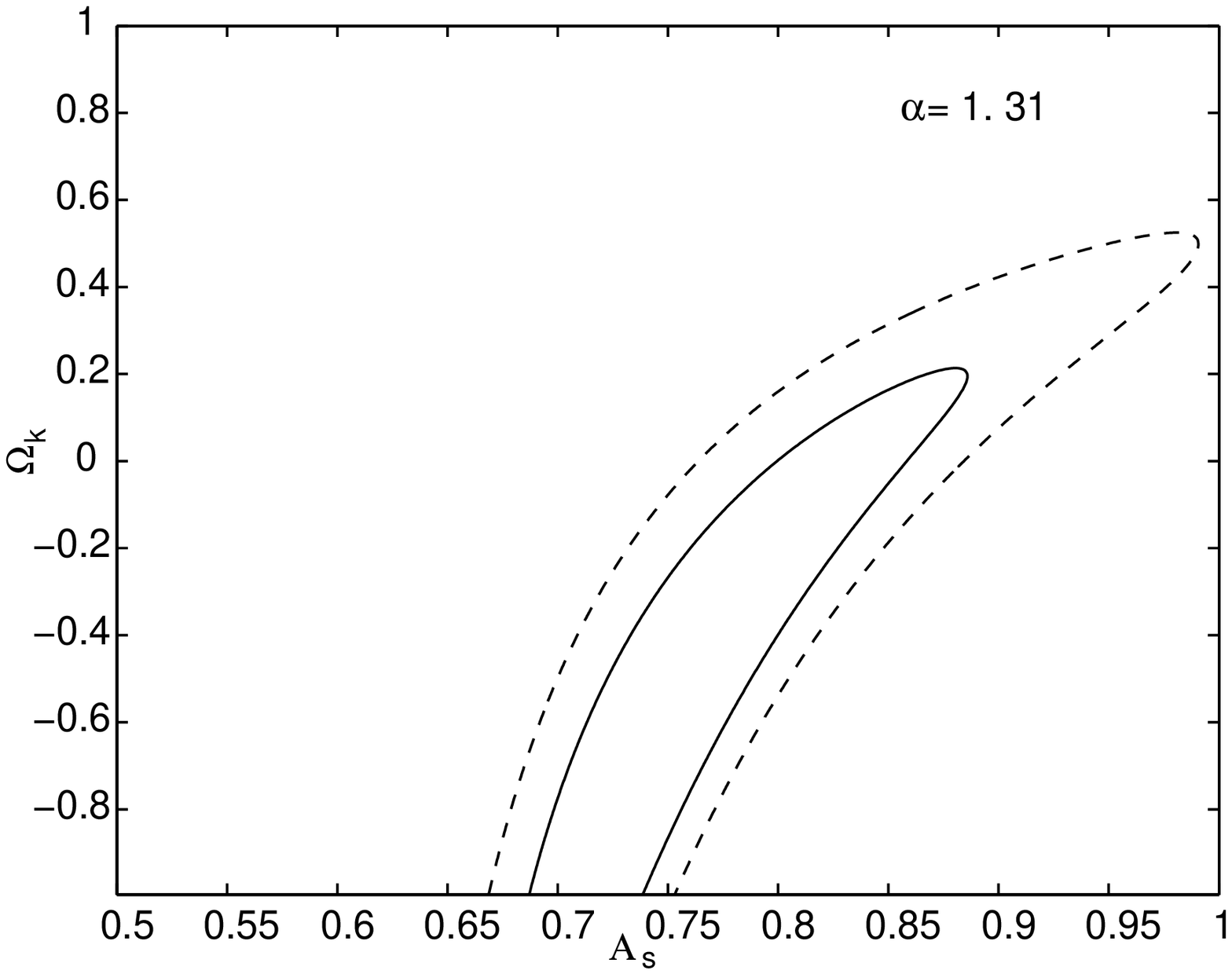}
 \includegraphics[height=6.5cm]{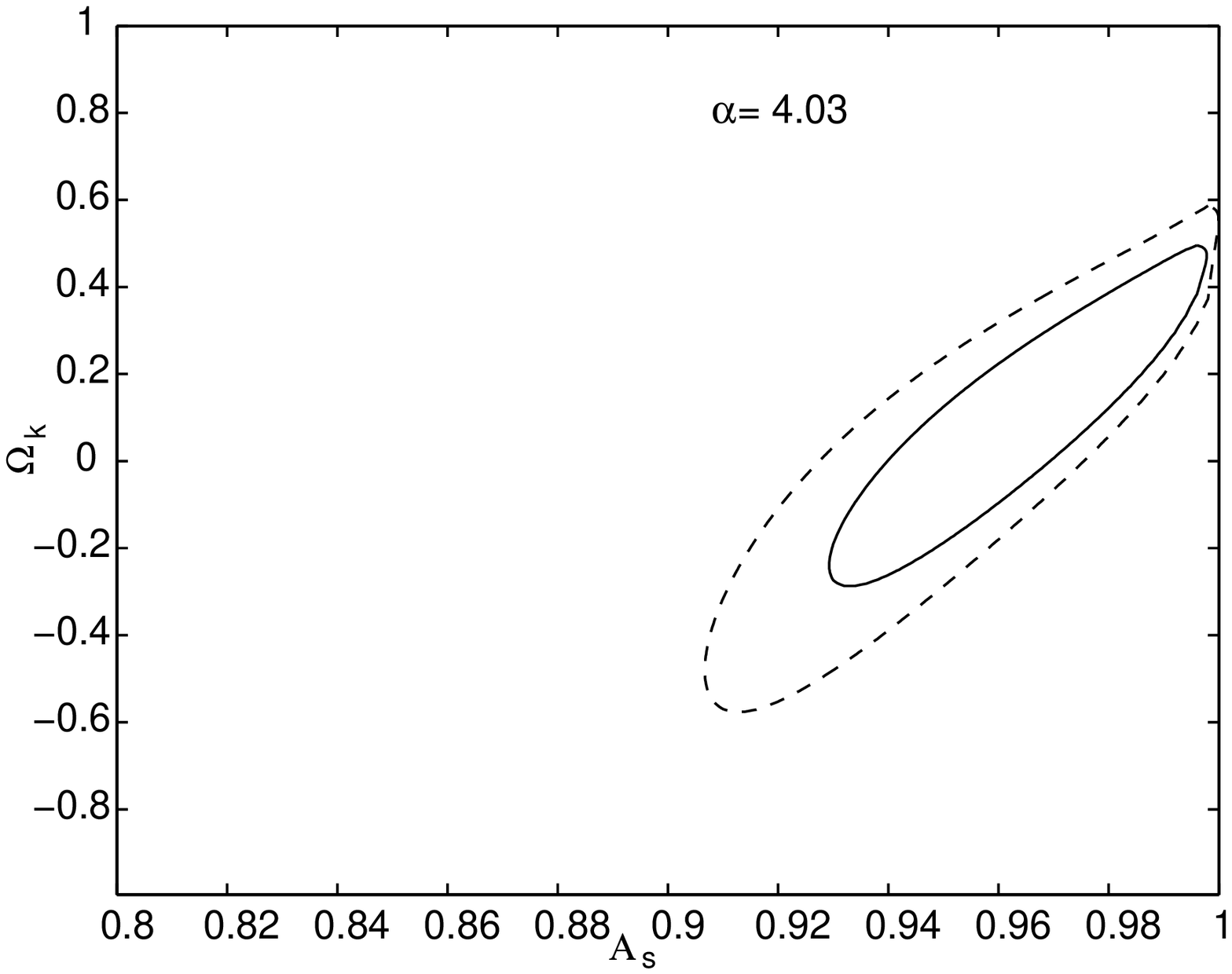}
 \includegraphics[height=6.5cm]{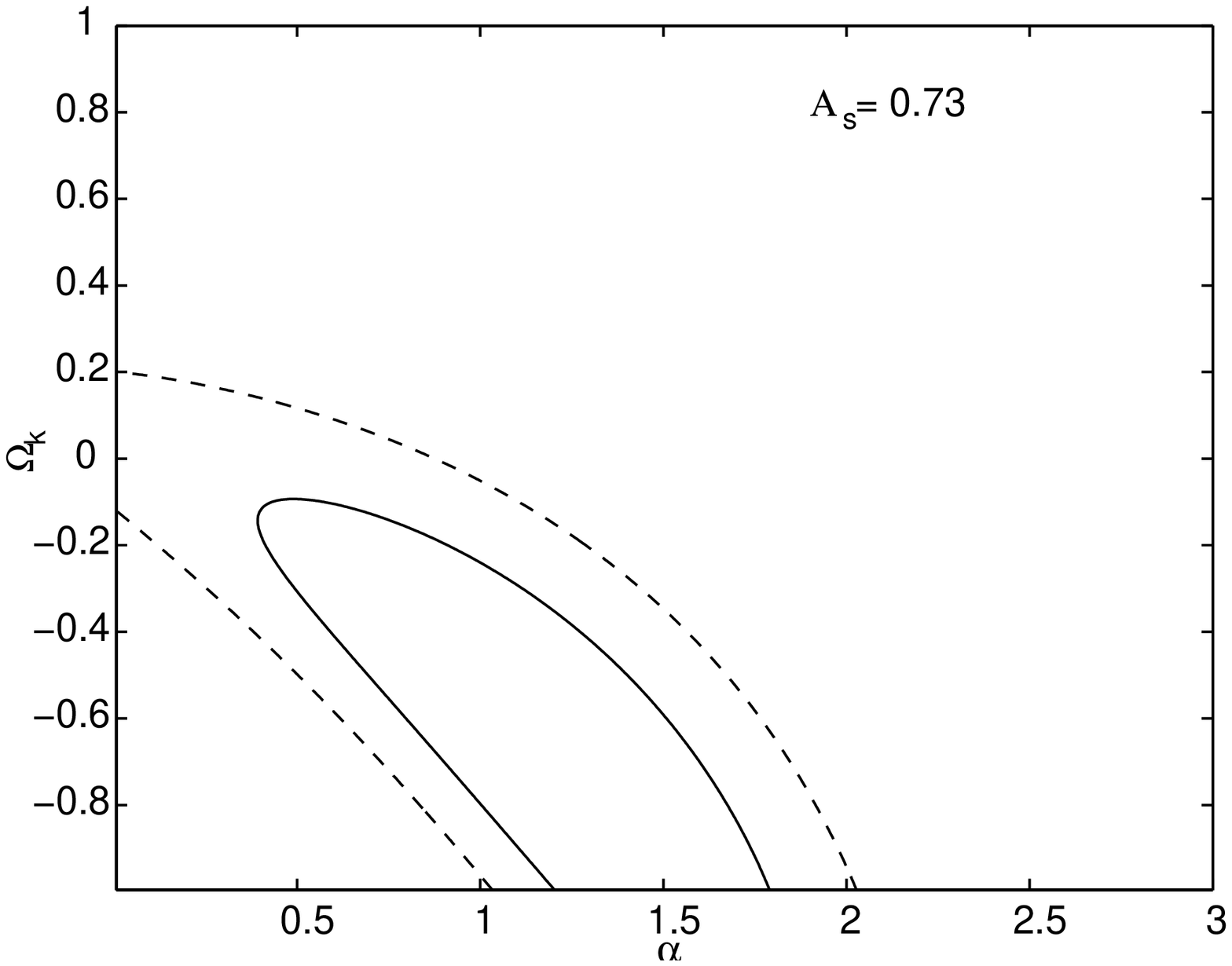}
 \includegraphics[height=6.5cm]{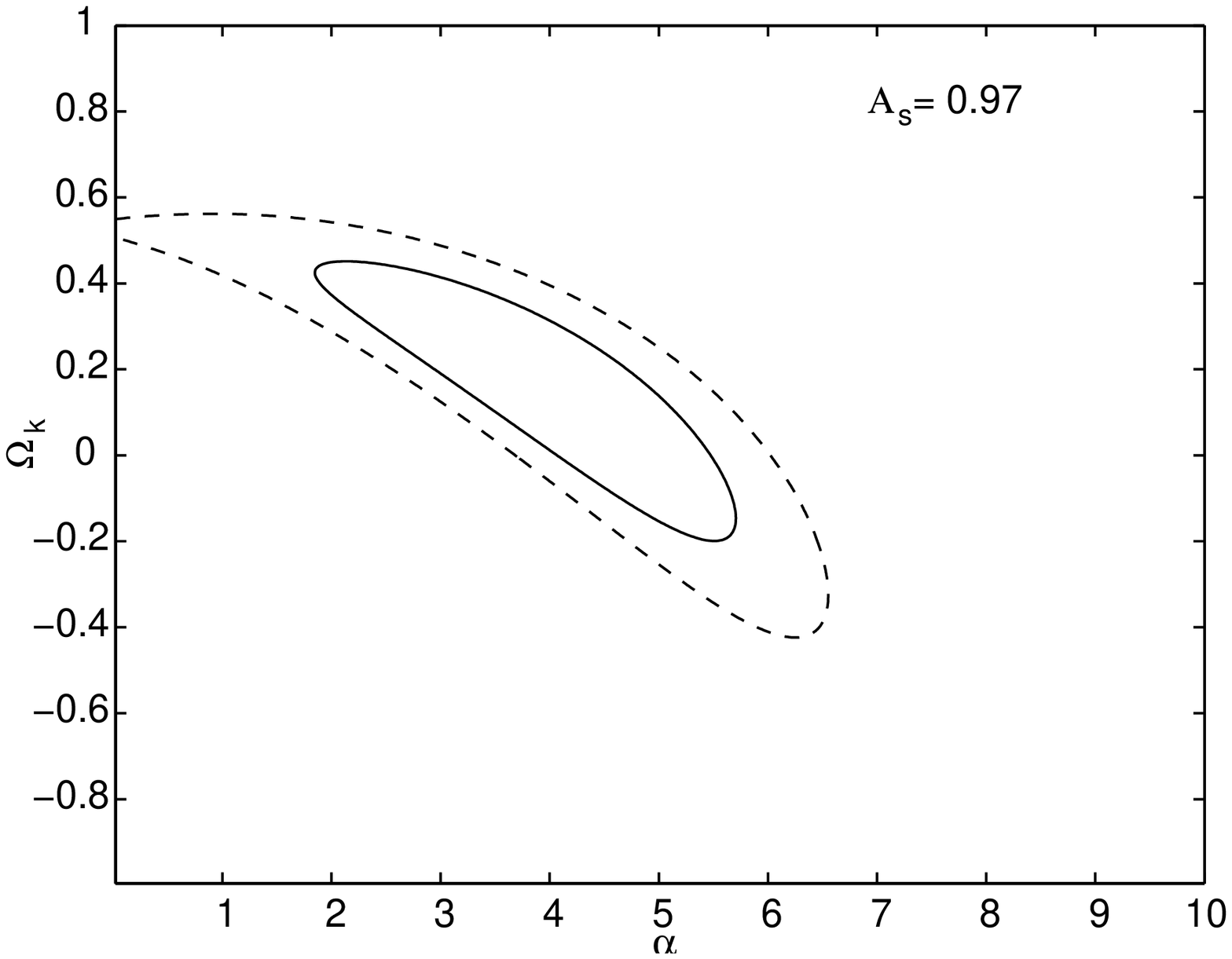}
  \caption{\label{fig:GCG} Confidence contours in the and
  $\Omega_k-A_s$ and $\Omega_k-\alpha$
parameter space for the non-flat GCG model. The
  solid and dashed lines represent
  the $68\%$ and $95\%$ confidence regions, respectively. }
\end{center}
\end{figure*}

The Friedmann equation for a non-flat unified GCG model in general
is given by
\begin{align}
\left({ {H} \over {H_0} }\right)^2  =
\left(1-\Omega_{k}\right)&\left[A_{s} +
(1-A_s)(1+  z)^{3(1+\alpha)}\right]^{1/(1+\alpha)} \nonumber\\
+ & \Omega_{k}(1+z)^{2}~~.\label{FriedGCGmodel}
\end{align}

This model has been thoroughly scrutinized from the observational
point of view; indeed, its compatibility with the Cosmic Microwave
Background Radiation (CMBR) peak location and amplitudes
\cite{Bento2003,Finelli}, with SNe Ia data
\cite{Supern,Bertolami2004} and gravitational lensing statistics
\cite{Silva,Alcaniz} has been analysed by various groups. The
issue of structure formation
\cite{Bilic:2001cg,Bento:2002ps,Fabris} and its dificulties
\cite{Sandvik} have been recently addressed \cite{Bento2004}. More
recent analysis, based on the 194 SNe Ia data points from Ref.
\cite{Tonry:2003zg}, has yielded interesting  results in what
concerns the allowed parameter space of the model
\cite{Bertolami2004}:

\vskip 0.2cm \noindent
1) Data favors $\alpha > 1$, although
there is a strong degeneracy on $\alpha$. At $68 \%$ confidence
level the minimal allowed values for $\alpha$ and $A_s$ are $0.78$
and $0.778$, which rules out the $\Lambda$CDM model $\alpha = 0$
case. However, at $95 \%$ confidence level it is found that there
is no constraint on $\alpha$.

\vskip 0.2cm
\noindent
2) Dropping the assumption of flat prior, it is found that GCG is
consistent with data for values of $\alpha$ sufficiently different
from zero. Allowing for some small curvature, positive or negative,
one finds that the GCG model is a more suitable description than the
$\Lambda$CDM model.

\vskip 0.2cm
\noindent
These results are similar to the ones obtained in
Refs. \cite{Alam:2003, Choudhury:2003tj}, where it is concluded that the supernova
data of Ref. \cite{Tonry:2003zg} favors ``phantom''-like matter with an
equation of state of the form $p = \omega \rho$ with $\omega < -1$.

In our present analysis, we have new results both with and without
flat prior. For the flat case, the $68\%$ and $95\%$ confidence
level contours are quite similar both with and without HST data
with that obtained by \cite{Bertolami2004} as shown in Fig.
\ref{fig:GCGflat}. But still there is one interesting feature to
note. It appears now that without the HST data, $\Lambda$CDM model
($\alpha = 0$) is ruled out even at $95\%$ confidence level which
was not the case in the previous analysis. But incorporating the
HST data, again makes $\Lambda$CDM model consistent at $95\%$ C.L
although it is still ruled out at $68\%$ C.L.

When we relax the condition of flat prior, one can have some
interesting features (see Fig. 6). Without the HST data, supernova
data prefers a nonflat model, but flat case is consistent both at
$68\%$ and $95\%$ C.L if one choose a value for $\alpha$
sufficiently different from zero. This means without flat prior,
and without HST data, data prefers a curved model for GCG but flat
case is still consistent which is similar to the conclusion in
\cite{Bertolami2004}.

But now if one includes the HST data, the best fit model itself
becomes a very close to flat case ($\Omega_k = 0.02$, Table
\ref{table:best}) which is completely a new feature. This shows
that with the new HST data, the current gold sample of supernova
data prefers a flat GCG model which is consistent with CMBR
observation. It also shows that GCG is a better choice among the
three possible choices of alternative models that we have
considered here as far the gold sample of supernova data including
those from HST measurements are concerned.

\begin{table*}[t]
\begin{tabular}{c c c c}
 \hline\hline
 Model & $q_0$ & \hspace{6mm} &
$\left.{dq\over{dz}}\right|_0$\\
\hline XCDM & ${1 \over 2} + {3 \over 2} w_X \Omega_X$ && ${9
\over 2 }w_X^2
\Omega_X (1-\Omega_X)$ \\
\hspace{4mm} Cardassian \hspace{4mm} & ${1 \over 2} + {3 \over
2}(n-1) (1-\Omega_{m})$ && ${9 \over 2 }(n-1)^2 \Omega_{m}
(1-\Omega_{m})$\\
DGP  & ${1 \over 2} + {3 \over 2}{ \Omega_{m}-1 \over 1 +
\Omega_{m}}$ && ${9 \Omega_m (1-\Omega_m) \over (1+\Omega_m)^3}$
\\  GCG &
${3\over{2}}(1-A_s) - 1$ && ${9\over{2}}A_s(1-A_s)(1+\alpha)$ \\
\hline \hline
\end{tabular}
\caption{Deceleration and jerk parameters for XCDM, Cardassian,
DGP and GCG models.} \label{table:statefinders}
\end{table*}

\section{Degeneracy with XCDM model}

To illustrate the degeneracy between our models and the XCDM
model, with constant equation of state $P/\rho=w_X$ and dark
energy density $\Omega_X=1-\Omega_{m}$, let us consider the Taylor
expansion of the luminosity distance as
\begin{align}
\label{dLseries} d_L = {c\over{H_0}}\left[ z +
{(1-q_0)\over{2}}z^2 -
  {1\over6}(1-q_0-3q_0^2+j_0)z^3 + ...\right]~,
\end{align}
where $q_0$ is the deceleration parameter, related with the second
derivative of the scale-factor, and $j_0$ is the so-callled statefinder or jerk
parameter \cite{Visser:2003vq},  related with the third derivative
of the scale-factor.  The subscript ``0'' means that quantities
are evaluated at present. The jerk parameter is related with the
deceleration parameter
 $q_0$ as
\begin{align}
j_0 = q_0 + 2q_0^2 + \left.{dq\over{dz}}\right|_0~.
\end{align}
Notice that the jerk paramter is  related to the geometry of the
Universe (see Ref. \cite{Caldwell}, where it is shown that the
measurement of the cubic correction to the Hubble law via
high-redshift supernovae is the first cosmological measurement,
after the CMBR, that probes directly the effects of spatial
curvature). We have calculated $q_0$ and
$\left.{dq\over{dz}}\right|_0$ for our models and obtained the
results shown in Table \ref{table:statefinders}. We have
considered only the flat case and, for simplicity, we study the
DGP model instead of the more generic DT model; $\Omega_m$ is the
only parameter of this model.

For low redshifts, it is sufficient to consider the first two
terms in the series expansion of the luminosity distance, Eq.
(\ref{dLseries}). From the expression of $q_0$ for the GCG, we can
see that in this case the SNe Ia can only constrain $A_s$, as
$q_0$ is independent of $\alpha$. However for the Cardassian model
the $q_0$ dependence is both on $\Omega_m$ and $n$, allowing low
redshifts SNe Ia to put constraints in the model.

Moreover, in order to have degeneracy between the models we are
analyzing and XCDM model, the $q_0$ parameter of these models must
be equal, $q_0^{XCDM}=q_0^{{\rm ~Model}~i}$, which results that:
\begin{align} w_X&=(n-1) {1-\Omega_m \over \Omega_X}~~~~~{\rm
for ~Cardassian
~model}~;\\
 w_X&=-{1- \Omega_{m} \over (1 + \Omega_{m}) \Omega_X}~~~~~{\rm for~DGP~model}~;\\
w_X&= - {A_s \over \Omega_X}~~~~~{\rm for ~GCG
~model}~\label{GCGwX}.
\end{align}

If one goes to higher redshifts, one also has to consider the
higher order terms in the expansion of $d_L(z)$; hence, in this
case also the jerk parameters have to be equal, which means that
$\left.{dq\over{dz}}\right|_0$ have to be the same. We get that
for Cardassian model
\begin{align}
w_X&=n-1~, \\
\Omega_X&=1-\Omega_m~,
\end{align}
meaning that the dynamical evolution of this model is equivalent
to a dark energy model with  the same matter density and a
constant equation of state given by
$w_{eff}=n-1$; the equivalent dark energy potential can be
written as $V(\phi)=A \left[\sinh k
(\phi/\sigma+C)\right]^{-\sigma}$ with $\sigma=-2-2/(n-1)$
\cite{Gong:2004sa} (see also Ref. \cite{Varun}).
We have seen that negative values of $n$ are
preferred which is consistent with the
fact that phantom equation of state, $w_X<-1$, is favored by the
data \cite{Alam:2003, Choudhury:2003tj, Bertolami2004}.

For DGP model one finds
\begin{align}
w_X&={\Omega_m^2-2 \Omega_m-1 \over (1+\Omega_m)^2}~,\\
 \Omega_X&={\Omega_m^2-1 \over \Omega_m^2-2 \Omega_m-1}~.
\end{align}
We see that for $\Omega_m<0.6$ the the equation of state is
phantom-like and, as $\Omega_m \rightarrow 0$, $w_X
\rightarrow-1$. We saw that the SNe Ia prefer $\Omega_m \sim 0.2$,
consistent with the fact that phantom equation of state is
favored. Moreover, the amount of matter needed in a XCDM model to
be equivalent to a given DGP model is larger,
$\Omega_m^{XCDM}=1-\Omega_X \geq \Omega_m$.

For the GCG model we find
\begin{align}
w_X&=- \alpha (1-A_s)-1~,\\
 \Omega_X&= {A_s \over 1+ \alpha (1-A_s)}~,
\end{align}
We see that for any GCG model, the corresponding dark energy model
equation of state has to be always phantom type
\cite{Bertolami2004}, as $\alpha > 0$ and $0 < A_s < 1$.
Nevertheless, for low redshift data, GCG is degenerate with all
kinds of constant equation of state dark energy model, including
$\Lambda$CDM model, as can be inferred from Eq. (\ref{GCGwX}).

\vskip 0.5cm

\section{Discussion and Conclusions}

In this paper we have performed likelihood analysis of the latest
type Ia supernova data for three distinct dark matter models. We
have considered the Cardassian model, the modified gravity
Dvali-Turner model and the generalized Chaplygin gas model of
unification of dark energy and dark matter. We find that SNe Ia
most recent data allows, in all cases, for non-trivial constraints
on model parameters as summarized in Table II.  We find that for
all models relaxing the flatness condition implies that data
favors then a considerable negative curvature for the gold without
HST dataset. For the gold dataset the resulting best fit value for
the curvature is positive (the GCG model is nearly flat in this
case). For all models we have found, in what concerns the
deceleration and jerk parameters, the conditions under which they
are degenerate to the XCDM model. Thus, SNe Ia data clearly favors
phantom-like equivalent equations of state.

Finally, in what concerns the gold sample of supernova data including
those from HST measurements, our analysis reveals that that the GCG flat
model is the better choice among the
three possible alternative models that we have
considered in this paper.

\begin{acknowledgments}
The work of N.M.C.S. and A.A.S was supported by Funda\c c\~ao para a
Ci\^encia e a Tecnologia (FCT, Portugal) under the grants
SFRH/BD/4797/2001 and SFRH/BPD/12365/2003, respectively.
\end{acknowledgments}

\end{document}